\documentstyle[preprint,prd,aps]{revtex}
\begin{document}
%\draft
%%%%%End of Preamble
%%%%Start of Text%%%%%%%%%%%%%%%%%%%%%%%%%%%%%%%%%%%%%%%%%%%%%%%%%%%%%%%
\tightenlines 
\preprint{
\vbox{\halign{&##\hfil\cr
		& OHSTPY-HEP-T-97-18  \cr
		& hep-ph/9710357      \cr
		& October 14, 1997    \cr
                &\vspace{0.6truein}   \cr
}}}
   
\title{ Renormalons in Electromagnetic Annihilation Decays of Quarkonium }

\author{Eric Braaten and Yu-Qi Chen}
\address{Physics Department, Ohio State University, Columbus, Ohio 43210}

\maketitle

\begin{abstract}
We study the large-order asymptotic behavior of the perturbation series
for short-distance coefficients in the NRQCD factorization formulas for 
the decays $J/\psi \to e^+e^-$ and $\eta_c \to \gamma\gamma$. The 
short-distance coefficients of the leading matrix elements  
are calculated to all orders  in the large-$N_f$ limit. We find that 
there is a universal Borel resummable renormalon associated with the
cancellation of the Coulomb singularity in the short-distance coefficients. 
We verify  that the ambiguities in the short-distance coefficients 
from the first infrared renormalon are canceled by ambiguities  
in the nonperturbative NRQCD matrix elements that contribute 
through relative order $v^2$.  Our results are used to estimate 
the coefficients of higher order radiative corrections in the decay 
rates for $J/\psi \to e^+ e^-$, $\eta_c \to \gamma\gamma$, and 
$J/\psi \to \gamma\gamma\gamma$. 
\end{abstract}


\vfill \eject

\narrowtext 

%%%%%%%%%%%%%%%%%%%%%%%%%%%%%%%%%%%%
\section{Introduction} 
 The annihilation decay  of heavy quarkonium involves
several  distance  scales.  The annihilation 
of the heavy-quark pair is a  short-distance process occurring at length 
scales of order $1/m$, where $m$ is the mass of the heavy quark. 
The probability that the heavy-quark pair will be close enough to annihilate
is determined by the wavefunction  of the heavy quarkonium, which is 
dominated by distances of order $1/mv$ or larger, where $v$ 
is the typical relative velocity of the heavy quark.  
The nonrelativistic QCD (NRQCD) factorization formalism\cite{B-B-L,Braaten} 
is a systematic framework for separating the effects of
the scale $m$ from effects of lower momentum scales. The 
annihilation decay rate is factored into short-distance coefficients 
and long-distance NRQCD matrix elements. The short-distance coefficients 
can be calculated using perturbative QCD as a power series in the strong 
coupling constant $\alpha_s(m)$ at the scale of the heavy quark mass. 
The matrix elements are nonperturbative, but scale in a definite way 
with $v$. The NRQCD factorization formalism therefore organizes the 
decay rate into a double expansion in powers of $\alpha_s$ and $v$.

In principle,  the short-distance coefficient for each matrix element 
can be calculated to any desired  order in $\alpha_s$ by matching  
the NRQCD effective theory with full QCD. One might hope that any desired 
precision could be obtained by carrying out the matching  calculations
to sufficiently high order in $\alpha_s$. However, large numbers 
may arise in  high order calculations through the factorial growth 
of the coefficients in the perturbation series.
A factorially growing series, when truncated at a 
high order, may produce a large error even if   
the coupling constant is small. This may 
limit the possibility of obtaining sufficiently precise results
by performing the perturbative calculations to  higher orders. 

One can investigate the asymptotic behavior of a series 
$f(\alpha_s)$ by studying its Borel transform $f(t)$.
A singularity  in the Borel parameter $t$ is called a 
renormalon\cite{thooft}. 
A renormalon gives rise to $n!$ growth of the
coefficients in the power series for $f(\alpha_s)$ so that its radius  of
convergence is 0.  If the renormalon is on the negative real axis, 
Borel resummation can be used to recover the function uniquely from the
divergent series.  If the  
renormalon is on the positive real axis, the sum of the divergent series is
not unique and there is an ambiguity in the function $f(\alpha_s)$. 

The renormalons that are responsible for the factorial growth of coefficients 
in the perturbation series for QCD have been extensively studied. Important
applications include the operator product expansion for $e^+e^-$ annihilation
and $\tau$ decay\cite{Mueller,Beneke,Neubert,BBB}
and heavy quark effective theory\cite{BB94,BSU,NS94,LMS94}. 
The large-$N_f$ limit of QCD has been developed by
Beneke\cite{Beneke} into a useful tool for the explicit study of renormalons. 
In this paper, we apply this tool to the NRQCD factorization formulas for 
$S$-wave quarkonium states, focusing on  electromagnetic 
annihilation decays.   

The rest of this paper is organized as follows. In section II, we 
calculate the short-distance coefficients for the electromagnetic
annihilation decays $J/\psi \to e^+ e^-$ and $\eta_c \to \gamma \gamma$
at tree level. We present a new method for calculating the tree-level 
short-distance coefficients which involves expanding nonlocal QCD
matrix elements in terms of local QCD matrix elements, and then using a  
Foldy-Wouthuysen-Tani  transformation to express them in terms of NRQCD matrix 
elements. In section III, we calculate the Borel 
transforms of the short-distance coefficients of the leading matrix elements 
in the large-$N_f$ limit. In section IV, we study the  
renormalons in the short-distance coefficients. 
We show that there is  a universal Borel  resummable  renormalon 
related to the cancellation of the 
Coulomb  singularity. We  demonstrate that the ambiguities from the 
first infrared renormalon are canceled by ambiguities in the
matrix elements that contribute through relative order $v^2$. 
We also deduce the leading renormalons for the decay $J/\psi \to
\gamma\gamma\gamma$.
Finally, in section V, we use our calculations to estimate higher order
radiative corrections to the decay rates for $J/\psi \to e^+e^-$, $\eta_c
\to \gamma\gamma$, and $J/\psi \to \gamma\gamma\gamma$.  

%%%%%%%%%%%%%%%%%%%%%%%%%%%%%%%%%%%%%%%%%%%%
\section{ short distance coefficients at tree level}
In this section, we review the NRQCD factorization formalism as it applies to
electromagnetic annihilation decays. We then calculate tree-level
short-distance coefficients for the decays $J/\psi \to e^+e^-$ and $\eta_c \to
\gamma\gamma$ and for several quarkonium-to-vacuum matrix elements.

%%%%%%%%%%%%%%
\subsection{ NRQCD Factorization Formalism} 
In the NRQCD factorization formalism, 
the annihilation decay rate for  the charmonium state $H$ 
is written in the factorized form \cite{B-B-L}
\begin{equation}
\Gamma ( H ) \;=\; {1 \over 2 M_{H} } \; 
\sum_{mn} {C_{mn} } \langle H | {\cal O}_{mn} | H \rangle \;,
\label{fact-Gam}
\end{equation}
where $M_H$ is the  mass of the  state $H$. The matrix elements 
$ \langle H | {\cal O}_{mn} | H \rangle $ are expectation values 
in the quarkonium state $H$ of local 4-fermion operators that 
have the structure
\begin{equation}
{\cal O}_{mn}
\;=\; \psi^\dagger {\cal K}_m \chi \;
	\chi^\dagger {\cal K}_n \psi \;,	
\label{O_mn}
\end{equation}
where $\psi$ and $\chi$ are the field operators for the heavy quark and 
antiquark in NRQCD, ${\cal K}_n$ and ${{\cal K}}_m$
are products of a color matrix ($1$ or $T^a$), 
a spin matrix ($1$ or $\sigma^i$), and a polynomial in the gauge-covariant 
derivative ${\bf D}$ and in the field strengths ${\bf E}$ and ${\bf B}$. 
It is convenient to define the states 
$ |H \rangle = | H ( {\rm\bf P } = 0 ) \rangle $
in (\ref{fact-Gam}) so that they have the 
standard relativistic normalization: 
\begin{equation}
\Big \langle H({\bf P}') \Big | H({\bf P}) \Big\rangle 
\;=\; 2 E_P \; (2 \pi)^3  \delta^3 ({\bf P} - {\bf P}') \;,	
\label{norm-H}
\end{equation}
where $E_P = \sqrt{M_H^2 + {\bf P}^2 }$. 
The coefficients $C_{mn}$ take into account the effects of 
short distances of order $1/m$, and they can therefore 
be calculated as perturbation series in the QCD coupling 
constant $\alpha_s(m)$.  

For electromagnetic annihilation decays, the factorization formula can be
simplified. Because there are no hadrons in the final state, one can insert a
projection $|0\rangle \langle 0 | $ onto the vacuum state between $\chi$ and
$\chi^\dagger$ in the operator (\ref{O_mn}). The factorization formula
(\ref{fact-Gam}) then becomes 
\begin{equation}
\Gamma ( H \to {\rm EM}  )
\;=\; 
{1 \over 2 M_{H} } \; 
\sum_{mn} {C_{mn} } 
     	\langle H | \psi^\dagger {\cal K}_m \chi 
     	|0\rangle \langle 0 |
	\chi^\dagger {\cal K}_n \psi | H \rangle \;,
\label{fact-Gam:EM}
\end{equation}
where EM represents an electromagnetic final state consisting of photons or
$e^+e^-$ pairs. The matrix elements vanish unless the operators are color 
singlets, so ${\cal K}_n$ and $ {\cal K}_m $ are products of a spin matrix 
(1 or $\sigma^i$) and a polynomial in ${\rm\bf D}$, ${\bf E}$, and ${\bf B}$. 
The short-distance coefficients $C_{mn}$ in (\ref{fact-Gam:EM}) include the
effects of integrating over the phase space of the electrons, positrons and
photons in the final state. The NRQCD factorization can actually be carried out
before the phase space integration at the level of the $T$-matrix element: 
\begin{equation}
{\cal T}_{H\to {\rm EM} } \;=\; \sum_m C_m \,\langle 0 |
\chi^\dagger {\cal K}_m  \psi | H \rangle \;.
\label{fact-T:EM}
\end{equation}

The velocity-scaling rules of NRQCD can be used to determine how the matrix
elements in (\ref{fact-T:EM}) scale with the typical relative velocity 
$v$ of the heavy quark. For decays of the S-wave states $J/\psi$ and $\eta_c$, 
the leading contributions are from  NRQCD matrix elements 
$\langle 0 |\chi^\dagger \mbox{\boldmath $\sigma$} \psi | J/\psi \rangle $
and  $\langle 0 |\chi^\dagger \psi | \eta_c \rangle $, which scale like
$v^{3/2}$. The next most important matrix elements are
$\langle 0 |\chi^\dagger {\rm\bf D}^2 \mbox{\boldmath $\sigma$} 
 \psi | J/\psi \rangle $
and $\langle 0 |\chi^\dagger {\rm\bf D}^2 \psi | \eta_c \rangle $,
which are suppressed by $v^2$ and therefore represent relativistic 
corrections. The matrix elements are related by the approximate 
heavy-quark spin symmetry of NRQCD: 
\begin{eqnarray}
\langle 0 |\chi^\dagger \mbox{\boldmath $\sigma$} \psi 
   | J/\psi (\mbox{\boldmath $\epsilon$}) \rangle 
 &\;=\; &  
 \mbox{\boldmath $\epsilon$} \;\langle 0 |\chi^\dagger  \psi | \eta_c \rangle \; 
 \left(1 +O(v^2) \right) \;,
 \label{spin-1}\\ 
\langle 0 |\chi^\dagger {\rm\bf D}^2 \mbox{\boldmath $\sigma$} \psi 
   | J/\psi (\mbox{\boldmath $\epsilon$}) \rangle 
 &\;=\; &  
 \mbox{\boldmath $\epsilon$} \; \langle 0 |\chi^\dagger {\rm\bf D}^2  
          \psi | \eta_c \rangle \; \left(1 +O(v^2) \right) \;,
 \label{spin-D2}
\end{eqnarray}  
where $\mbox{\boldmath $\epsilon$} $ is the polarization vector of the $J/\psi$
which satisfies $ \mbox{\boldmath $\epsilon$}^2 = 1 $.  
Thus there are only 3 independent matrix elements that contribute to $S$-wave
decays through relative  order  $v^2$.

The short-distance coefficients in the factorization formula (\ref{fact-T:EM})
can be determined by the covariant projection method\cite{Mang} or by the
threshold expansion method\cite{Braaten-Chen}. In the threshold expansion
method, the coefficients are determined 
by matching perturbative $T$-matrix elements for $c\bar{c}$
annihilation. 
Let $|c\bar{c}\rangle \equiv | c({\bf q},\xi) \bar c(-{\bf q},\eta) \rangle$ 
represent a state 
that consists of a $c$ and a $\bar c$ with spatial momenta $ \pm {\bf q}$ 
in the charmonium rest frame and with spin and color states specified by  
spinors $\xi$ and $\eta$ that are normalized to 
$\xi^\dagger \xi = \eta^\dagger \eta = 1$. 
The standard relativistic normalization is
\begin{equation}
\Big \langle c({\bf q}_1',\xi') \bar c({\bf q}_2',\eta')
	 \Big | c({\bf q_1},\xi) \bar c({\bf q_2},\eta)  \Big \rangle 
\;=\; 4 E_{q_1} E_{q_2} \; (2 \pi)^{6}  \delta^3 ({\bf q}_1 - {\bf q}_1')
	\delta^3({\bf q}_2 - {\bf q}_2') \; 
	\xi'^\dagger \xi {\eta'}^\dagger \eta \; ,
\label{cc-norm}
\end{equation}
where $E_q = \sqrt{m^2 + {\bf q}^2}$.  
In the threshold expansion method, the matching condition is 
\begin{eqnarray}
&& {\cal T}_{ c \bar c \to  {\rm EM} } \Big|_{pQCD}
  \;=\; \sum_{m} {C_{m} } \;
	\langle 0 |\chi^\dagger {\cal K}_m \psi | c \bar{c} \rangle 
	\Big|_{pNRQCD} \;,
\label{match}
\end{eqnarray}
where  ${\cal T}_{ c \bar c \to  {\rm EM} }$ is the $T$-matrix element 
for the annihilation of the  $c \bar{c}$ into the electromagnetic final state EM. 
To carry out the matching procedure, the left side of (\ref{match}) is
calculated using perturbation theory in full QCD, and then expanded 
in powers of ${\bf q}$. The matrix elements on the right 
side of (\ref{match}) are 
calculated using perturbation theory in NRQCD, and then expanded 
in powers of ${\bf q}$. The short-distance coefficients
$C_{m}$ are obtained by matching the terms in the expansions
in ${\bf q}$ order by order in $\alpha_s$. 
To determine all the coefficients $C_m$, it is also necessary to consider the
annihilation of $c\bar{c}$ states with general momenta ${\bf q}_1$  and
${\bf q}_2$ and the annihilation of higher Fock states like 
$|c\bar{c}g \rangle$.
At tree level, some of these coefficients can be determined more easily 
using an alternative method that involves a
Foldy-Wouthuysen-Tani transformation. This method will be presented in the
next subsections.  

%%%%%%%%%%%%%%% 
\subsection {Leptonic Decay of $J/\psi $  } 
The decay $J/\psi \to e^+e^-$ proceeds through the annihilation of 
the $J/\psi$ into a virtual photon. 
The Feynman diagram for $c \bar{c} \to e^+ e^- $ is shown in Fig.~1.
The $T$-matrix element for the decay of $J/\psi$  into an electron and 
a positron with momenta $k_1$ and $k_2$ can be written as
\begin{equation}
{\cal T}_{J/\psi \to e^+ e^-} %(k_1,k_2) 
\;=\; -\,{Q e^2 \over M_\psi^2 } \; 
 {\bar{u}(k_1) \gamma_\mu v(k_2)  } \; 
 \langle 0 |  \bar{\Psi} \gamma^\mu \Psi | J/\psi \rangle \;,
\label{T-psiee:2}
\end{equation}
where $Q = + 2/3 $ is the electric charge of the charm quark
and $M_\psi$ is the mass of the $J/\psi$.
The factor of $1/M_\psi^2$ comes from the propagator of the virtual photon. 
The matrix element in (\ref{T-psiee:2}) is that of a local QCD operator 
between the $J/\psi$ and the vacuum state.
Here and below, the argument of a local operator is assumed to be 
$x=0$ if it is not given explicitly. 
We obtain the decay width by squaring the $T$-matrix element, summing
over the spins of the $e^+$ and $e^-$,   and 
integrating over the phase space of the $e^+e^-$ pair: 
\begin{equation}
{\Gamma} ( {J/\psi \to e^+ e^-} ) \;= \;
 { 4 \pi Q^2  \alpha^2 \over 3 M_{\psi}^3 } \; 
 (-g_{\mu\nu}) \,\langle J/\psi | \bar{\Psi} \gamma^\nu \Psi | 0 \rangle \;
 \langle 0 | \bar{\Psi} \gamma^\mu \Psi | J/\psi \rangle \;.
\label{Wid-psiee:2}
\end{equation}

The QCD matrix  element  
$\langle 0 | \bar{\Psi} \gamma^\mu \Psi | J/\psi \rangle $ in 
(\ref{Wid-psiee:2}) 
includes effects of both short distances of order $1/m$ or smaller 
and long distances of order $1/mv$ or larger. 
We can separate these effects by using the NRQCD factorization formalism 
to express  the QCD  matrix element as a sum of NRQCD matrix elements 
multiplied by short-distance coefficients.
At tree level, this is most easily accomplished by a 
Foldy-Wouthuysen-Tani (FWT) transformation\cite{F-W-T}. This transformation 
can be constructed order by order in $ v$ so that off-diagonal terms in the 
Hamiltonian that couple the heavy quark and antiquark are suppressed 
to any desired order in $v$. Up to corrections of order 
$v^3$, the FWT transformation is  
\begin{eqnarray}
 \Psi(x) & \; = \; &  \exp 
 \left( { {i \over 2 m } \mbox{\boldmath $\gamma$} \cdot {\rm\bf  D}} \right) \;
 \left( 
 \begin{array}{c} 
  \psi(x) \\ 
  \chi(x) 
 \end{array} \right) \;,
\label{2-4}
\end{eqnarray}
where ${\bf D}= \mbox{\boldmath $\nabla$} - i g_s {\bf A} $ is the 
covariant derivative. Inserting (\ref{2-4}) into the matrix element 
$\langle 0 |  \bar{\Psi} \gamma^\mu \Psi | J/\psi \rangle$, expanding to
second order in ${\rm\bf D}/m$, and keeping only the annihilation terms that
involve $\chi^\dagger$ and $\psi$, we obtain
\begin{eqnarray}
 && \langle 0 | \bar{\Psi} \gamma^\mu \Psi | J/\psi \rangle 
  =\; -g^{\mu i} \;\Bigg[\;
  \langle 0 | \chi^\dagger \sigma^i \psi | J/\psi \rangle
  \;+\; {1 \over 4m^2}  
  \langle 0 | \chi^\dagger 
       ( D_i D_j + D_j D_i )
 \sigma^j  \psi |J/\psi \rangle\;\Bigg] \;.
\label{A-expan}
\end{eqnarray}
Up to this point, the only property of the $J/\psi$ that we have used is  
that it has the quantum numbers $J^{PC} =1^{--}$ that allow it 
to annihilate into a virtual photon. According to the velocity-scaling 
rules of NRQCD, the first term on the right side of (\ref{A-expan}) 
scales like $v^{3/2}$. 
The last term in (\ref{A-expan}) is a symmetric  tensor in the 
indices of the two covariant derivatives. 
The trace of this tensor contributes   at relative order $v^2$.
The traceless part  receives a contribution from 
the $D$-wave component of the 
$|c\bar{c} \rangle $ Fock state of the $J/\psi$ and is suppressed by $v^4$.
Keeping only the terms in (\ref{A-expan})  that contribute through relative 
order $v^2$, this matrix element reads  
\begin{eqnarray}
 && \langle 0 | \bar{\Psi} \gamma^\mu \Psi | J/\psi \rangle 
  =\; -g^{\mu i} \;\Bigg[\;
  \langle 0 | \chi^\dagger \sigma^i \psi | J/\psi \rangle
  \;+\; {1 \over 6m^2}  
  \langle 0 | \chi^\dagger {\bf D}^2 \sigma^i \psi |J/\psi \rangle  
  \;\Bigg]\;.
\label{A-expan:2}
\end{eqnarray}
Inserting (\ref{A-expan:2}) into  (\ref{Wid-psiee:2}), the decay 
width can be written as 
\begin{eqnarray}
{\Gamma}(J/\psi \to e^+e^-) &\;=\;& { 4 \pi Q^2  \alpha^2 \over 3 M_{\psi}^3 } \; 
 \bigg| \; 
   \langle 0 | \chi^\dagger \mbox{\boldmath $\sigma$} \psi | J/\psi \rangle
   \;+\; {1 \over 6m^2}  
  \,
   \langle 0 | \chi^\dagger {\bf D}^2 \mbox{\boldmath $\sigma$} \psi |J/\psi \rangle  
 \; \bigg|^2 \;.
\label{Wid-psiee:3}
\end{eqnarray}
This is the NRQCD factorization formula for the  decay width, including 
the first relativistic corrections and with the short-distance coefficients 
calculated to tree level. 

In Ref.\cite{B-B-L}, the factorization formula for the decay rate of $J/\psi
\to e^+ e^-$ was derived by matching $c\bar{c}$ scattering amplitudes. The
$T$-matrix element for $c\bar{c} \to c \bar{c}$ through an intermediate
$e^+e^-$ state was matched with the $T$-matrix element for $c\bar{c} \to c
\bar{c} $ through operators of the form 
$\psi^\dagger K_m \chi |0 \rangle \langle 0 | \chi^\dagger K_n \psi $ 
in NRQCD. Using our relativistic normalization of states and keeping 
only terms through relative order $v^2$, the result is  
\begin{eqnarray}
{\Gamma}(J/\psi \to e^+e^-) &\;=\;& 
{ \pi Q^2  \alpha^2 \over 3 M_{\psi}\, m^2  } \; 
 \bigg| \; 
   \langle 0 | \chi^\dagger \mbox{\boldmath $\sigma$} \psi | J/\psi \rangle
 \,+\, {2 \over 3 m^2} \, 
   \langle 0 | \chi^\dagger {\bf D}^2 \mbox{\boldmath $\sigma$} 
   \psi |J/\psi \rangle \, 
  \bigg|^2  \;.
\label{Wid-psiee:3p}
\end{eqnarray}
As pointed out by Maksymyk\cite{Maksymyk}, the consistency of (\ref{Wid-psiee:3}) and (\ref{Wid-psiee:3p}) through
relative order $v^2$ requires the  identity
\begin{equation}
   \langle 0 | \chi^\dagger {\bf D}^2 \mbox{\boldmath $\sigma$} 
   \psi |J/\psi \rangle  
  \;=\; -m (M_\psi - 2 m)\,  
   \langle 0 | \chi^\dagger \mbox{\boldmath $\sigma$} \psi | J/\psi \rangle \,
   \left(\, 1 + O(v^2) \, \right) \;.
\label{Gremm-Kapustin}
\end{equation}
This identity was recently derived by Gremm and
Kapustin\cite{Gremm-Kapustin}, and follows simply from the equations of
motion of NRQCD. It allows factors of $m$ in a factorization formula to be
traded for factors of $M_\psi$:
\begin{equation}
{M_\psi \over 2 m} \;=\; 1 \,-\, { 1\over 2 m^2}\,  
 { \mbox{\boldmath $\epsilon$} \cdot \langle 0 | 
 \chi^\dagger {\bf D}^2 \mbox{\boldmath $\sigma$} 
 \psi |J/\psi (\mbox{\boldmath $\epsilon$}) \rangle  
  \over
   \mbox{\boldmath $\epsilon$} \cdot \langle 0 | 
   \chi^\dagger \mbox{\boldmath $\sigma$} \psi 
   | J/\psi(\mbox{\boldmath $\epsilon$}) \rangle } \,
   +\,  O(v^4)  \;.
\label{Mpsi-m}
\end{equation}

The identity (\ref{Mpsi-m})  can also be used to relate our NRQCD matrix 
elements, which are defined using the standard relativistic 
normalizations for the quarkonium states, to the matrix elements of
Ref.\cite{B-B-L}, which were defined using the standard 
nonrelativistic normalizations:
\begin{equation}
\langle 0 | \chi^\dagger {\cal K}_n \psi | H \rangle 
\;=\; \sqrt{2M_H} \, 
\langle 0 | \chi^\dagger {\cal K}_n \psi | H \rangle_{\rm\small BBL} \;.
\label{norm:bbl}
\end{equation}
Using (\ref{Mpsi-m}) and (\ref{norm:bbl}), we obtain the relation 
\begin{eqnarray}
   \langle 0 | \chi^\dagger \mbox{\boldmath $\sigma$}  \psi | J/\psi \rangle 
&\;=\;& 
\sqrt{4 m} \,\bigg(\, 
\langle 0 | \chi^\dagger\mbox{\boldmath $\sigma$}   \psi |J/\psi 
\rangle_{\rm\small BBL} 
 \nonumber\\
& & \hspace{0.5in} 
\, -\,  {1\over 4 m^2}\, 
\langle 0 | \chi^\dagger {\bf D}^2 \mbox{\boldmath $\sigma$}  \psi |J/\psi 
\rangle_{\rm\small BBL} \, \bigg) \;
\left (\, 1 + O(v^4)\,\right) \;. 
\label{V-BBL}
\end{eqnarray}
Inserting  (\ref{V-BBL}) into  our factorization
formula (\ref{Wid-psiee:3}) and using (\ref{Mpsi-m}) to eliminate $M_\psi$ in
favor of $m$, we recover the result given in 
Ref.\cite{B-B-L}   up to corrections that are higher order in $v$: 
\begin{eqnarray}
{\Gamma}(J/\psi \to e^+e^-) &\;=\;& 
{ 2 \pi Q^2  \alpha^2 \over 3 m^2  } \; 
 \bigg| \; 
   \langle 0 | \chi^\dagger \mbox{\boldmath $\sigma$} \psi | J/\psi 
   \rangle_{\rm\small BBL}
 \,+\, {2 \over 3 m^2} \, 
   \langle 0 | \chi^\dagger {\bf D}^2 \mbox{\boldmath $\sigma$} 
   \psi |J/\psi \rangle_{\rm\small BBL} \, 
  \bigg|^2  \;.
\label{Wid-psiee:4}
\end{eqnarray}
%

%%%%%%%%%%%%%%%%%%%
\subsection { Decay of $\eta_c$ into  Photons  } 
If a quarkonium state $H$ is even under charge conjugation, it can decay into
two photons at  second order in the
electromagnetic interaction. The Feynman diagrams for $c\bar{c} \to
\gamma\gamma$ are shown in Fig.~2. 
If the photons have momenta $k_1$ and $k_2$ and polarization vectors 
$\mbox{\boldmath$\epsilon$}_1$ and $\mbox{\boldmath$\epsilon$}_2$, the
$T$-matrix element for the decay is
\begin{eqnarray}
{\cal T}_{H \to \gamma\gamma}  &\;=\;& - 16 Q^2 e^2  \; 
 {\epsilon_{1 \mu}   \epsilon_{2 \nu} }
 \int d^4 z \; e^{{i} ( k_1 -k_2 ) \cdot z } \nonumber \\
& & %\hspace{-0.2in} 
  \times \,  
     \langle 0 | \,\left[\, \bar{\Psi}\left( {z  } \right) 
     \, \gamma^\mu \,S \left( {z  }, -{z  } \right) \,
           \gamma^\nu 
    \Psi \left( - {z } \right)  
 \; + \;
      \bar{\Psi}\left( -{z  } \right) 
     \, \gamma^\nu \,S \left( -{z  }, {z  } \right) \,
           \gamma^\mu 
     \Psi \left(  {z  } \right) \,\right]\,| H \rangle    
  \;.
\label{A-etacgg}
\end{eqnarray}
The factor of 16 arises from taking the integration variable $z$ to be half 
the separation of the interaction points. 
This expression involves the full propagator $S (x,y)$ of the heavy quark, 
which satisfies the equation 
\begin{equation}
( i\not \!\! {D} -m )_x \; S (x,y) = \delta^4 (x-y) \;,
\label{pro-quark} 
\end{equation}
where $ D_\mu = \partial _\mu + i g_s A_\mu $. 

The phase factor in (\ref{A-etacgg}) and the behavior of the quark propagator
combine to give the integral over $z$ support only in the region where the components of
$z^\mu$ are of order of $1/m$. Since this is much smaller than the size 
$1/mv$ of the quarkonium, the operator in (\ref{A-etacgg}) can be expanded 
in powers of $z^\mu$. It is  convenient to carry out this expansion using  
the radial gauge\cite{radial}: 
\begin{equation}
z^\mu A_\mu(z) =0 \;.
\end{equation}
By Taylor expanding this gauge condition around $z=0$, one can show that
derivatives of $A_\mu$ at the origin can be expressed in terms of 
gauge-invariant operators. In particular, we have  $A_\mu(0)=0$ and  
$\partial_\mu A_\nu(0)=  F_{\mu\nu}(0)/2 $. 
The quark field in the radial gauge can be expanded as
\begin{equation}
\Psi\left( {z } \right) \;=\; \Psi(0)  
  \,+\, {z^\mu}\, D_\mu \,\Psi(0)  
  \,+\,{1\over 2}\, {z^\mu}\, {z^\nu}\, D_\mu \, D_\nu \,\Psi(0)
  \,+\,\cdots \;. 
\label{expan:field}
\end{equation}  
By solving (\ref{pro-quark}), we find that the quark propagator 
can be expanded as 
\begin{eqnarray}
  S \left( {z} ,-{z} \right)
  & \;=\;& S_F(2 z) \,+\, \int d^4 x\, 
  S_F \left({z} - x  \right) \, g_s \not \! \! {A}(x) \,
  S_F \left( x + {z} \right) \,+\, \cdots \nonumber \\
  & \;=\;& 
 \int_p\,  e^{ - 2ip \cdot z } \;
\Bigg[ \; { 1 \over \not \! {p} - m } \,-\, 
     {i\over 2} g_s F_{\mu\nu} (0)\, {1 \over (p^2 - m^2)^2 } 
     \gamma^\mu (\not\!{p} - m) \gamma^\nu \,+\, \cdots \;
\Bigg] \;, 
\label{expan:prop}
\end{eqnarray}
where $S_F ( x ) =\int_p \;{ e^{-ip\cdot x} / (\not \! {p} -m } ) $ 
is the Feynman  propagator for a  free quark. We have introduced  
an abbreviated notation for the integral over a four-momentum: 
\begin{equation}
\int_p \equiv \int \, {d^4 p \over (2\pi)^4} \;.
\end{equation}
Inserting (\ref{expan:field}) and (\ref{expan:prop}) into 
(\ref{A-etacgg}), expanding up to operators of dimension 5, and 
integrating over $z$, we obtain
\begin{eqnarray}
{\cal T}_{H \to \gamma\gamma} &\;=\;& - Q^2 e^2 \; 
 {\epsilon_{1 \mu}  \epsilon_{2 \nu}  } \; %\nonumber \\& & %\hspace{-0.7in}
% \times 
  \langle 0 | \bar{\Psi} \, \Bigg\{ \;
    {1\over r^2 - m^2}    
       \left( r_\lambda \, \Gamma_-^{\mu\lambda\nu} + 2 m g^{\mu\nu} \right)
\nonumber \\ & & \hspace{1.in}
    \;+\;i \, {\partial  \over \partial r_\alpha} \,
          {1 \over r^2 - m^2 } \, 
        \left( r_\lambda \, \Gamma_+^{\mu\lambda\nu} +  
                  m \Gamma_-^{\mu\nu} \right) \, D_\alpha 
\nonumber \\ & &\hspace{1in}
    \;-\; { 1\over 2 }\,{\partial  \over \partial r_\alpha} \,
                        {\partial  \over \partial r_\beta} \,    
                        {1 \over r^2 -m^2 } \, 
     \left( r_\lambda \, \Gamma_-^{\mu\lambda\nu} + 2 m g^{\mu\nu} \right)
     D_\alpha D_\beta 
\nonumber \\
& & \hspace{1in}
  \;-\; {1\over 2 } \, {1 \over (r^2 - m^2)^2 } \, 
  \left( \,
       r_{\lambda}\, \Gamma_+^{\mu\alpha\lambda\beta\nu} \, 
           - m \Gamma_-^{\mu\alpha\beta\nu} \, \right)\, 
        i g_s\, F_{\alpha\beta} \;
\Bigg\} \, \Psi  | H \rangle \; .
\label{expan-A-etac}
\end{eqnarray}
We can set $ r = (k_1 -k_2) /2 $ and $ r^2 = - M_{H}^2/4 $ after taking the
partial derivatives. We have used the notation 
\begin{equation}
\Gamma_\pm ^{\mu_1\mu_2\cdots \mu_N}  \;=\; 
\gamma^{\mu_1} \, \gamma^{\mu_2} \cdots \gamma^{\mu_N} \,\pm\,
\gamma^{\mu_N} \cdots \gamma^{\mu_2} \, \gamma^{\mu_1} 
\;.
\end{equation} 
To further simplify the $T$-matrix element, it is convenient to choose the photon
polarization vectors to be space-like in the rest frame of the $H$. In that
frame, we have $r_0=0$, ${\rm\bf r} \cdot \mbox{\boldmath$\epsilon$}_1 =0 $, 
and ${\rm\bf r} \cdot \mbox{\boldmath$\epsilon$}_2=0 $. The FWT transformation
(\ref{2-4}) can be used to reduce ${\cal T}$ to a sum of NRQCD matrix elements.
Keeping all terms through second order in ${\rm\bf D}/m$, we obtain  
\begin{eqnarray}    
{\cal T}_{H \to \gamma\gamma} &\;=\;& 
   -i\,{ 8 Q^2 e^2 \over M_H^2 + 4 m^2 } \; 
 { \epsilon^i_1 \, \epsilon^j_2  } \; 
 \langle 0 |  \chi^\dagger \, \Bigg\{\, 
   \epsilon^{ijk} r^k 
  \,+\, D_i \sigma^j + D_j \sigma^i  
  \,+\,{ 8 \, \delta^{ij} r^m r^n \over M_{H}^2 + 4m^2 } \,
   D_m \sigma^n \, \nonumber \\ 
 && \hspace{1.1in} 
     \;+\,{(M_{H}^2 + 12 m^2)\,\epsilon^{ijk} r^k    
           \over 2m^2( M_{H}^2 + 4 m^2) } \, 
     {\rm\bf D}^2  \, 
  \,-\,  
   { 64\, \epsilon^{ijk} r^k  r^m r^n \over (M_{H}^2 + 4 m^2)^2 }  \, 
 D_m D_n 
 \nonumber \\ 
 &&  \hspace{1.1in} 
     \;+\,{(M_{H}^2 - 4 m^2)\,\epsilon^{ijk} r^k    
           \over 2m^2( M_{H}^2 + 4 m^2) } \, 
   g_s {\rm\bf B} \cdot \mbox{\boldmath $\sigma$} 
 \nonumber \\      
 &&  \hspace{1.1in} 
  \,-\,    { 4m\,   \over M_{H}^2 + 4 m^2 }  \, 
    g_s (
      E^i \sigma^j +  E^j\sigma^i - \delta^{ij} 
        {\bf E} \cdot \mbox{\boldmath $\sigma$} ) \,
 \Bigg\} \, \psi | H \rangle 
\;. 
\label{expan-A-etac:2} 
\end{eqnarray} 
To determine the coefficients of the matrix elements involving ${\bf B}$ and 
${\bf E}$ using the threshold expansion method, we would have to match
scattering amplitudes for $c\bar{c}g \to \gamma\gamma$. 

Up to this point, the only property of the quarkonium state $H$ that 
we have used is that its charge conjugation quantum number is $C=+1$, 
so that it can decay into two photons. Thus the formula 
(\ref{expan-A-etac:2}) applies equally well to the $\eta_c$, the 
$\chi_{c0}$, and the $\chi_{c2}$. We now specialize to the $\eta_c$, 
which has quantum numbers $J^{PC} = 0^{-+} $. 
The parity conservation of NRQCD implies that the matrix elements of
operators linear in ${\bf D}$ or linear in ${\bf E}$ are zero. 
Using  the rotational invariance  of NRQCD, the remaining  matrix 
elements in (\ref{expan-A-etac:2}) can be reduced to scalars: 
\begin{eqnarray}         
{\cal T}_{ \eta_c \to \gamma\gamma } &\;=\;&  
 -i\, { 4 Q^2 e^2 M_{\eta_c} \over M_{\eta_c}^2 + 4 m^2 } \; 
 \mbox{\boldmath $\epsilon$}_1 \times \mbox{\boldmath $\epsilon$}_2 
 \cdot  {\bf \hat{r} }\,  
\nonumber \\ 
 && \hspace{0.2in} 
 \times\;\Bigg\{\, 
  \langle 0 | \chi^\dagger \psi | \eta_c \rangle 
     \; + \,
     {3 M_{\eta_c}^4 + 16  m^2 M_{\eta_c}^2 + 144 m^4 
      \over  6m^2(M_{\eta_c}^2 + 4 m^2)^2 } \; 
    %\right)\, 
     \langle 0 | \chi^\dagger \,{\rm\bf D}^2  \,\psi | \eta_c\rangle 
\nonumber \\ 
 && \hspace{1.0in} 
 \;+ \,
   {M_{\eta_c}^2 - 4 m^2 \over  2 m^2 (M_{\eta_c}^2 + 4 m^2) }\;
  \langle 0 |  \chi^\dagger  
  g_s {\rm\bf B} \cdot \mbox{\boldmath $\sigma$} \psi | \eta_c \rangle \,
 \Bigg\} \;.
\label{expan-A-etac:3} 
\end{eqnarray} 
According to the velocity-scaling rules, the matrix element 
$\langle 0 | \chi^\dagger \psi | \eta_c\rangle $ scales like $v^{3/2}$. 
The matrix element $\langle 0 | \chi^\dagger {\bf D}^2 \psi | \eta_c\rangle $
represents a relativistic correction that is suppressed by a power of $v^2$.  
The matrix element $\langle 0 | \chi^\dagger g_s {\bf B} \cdot 
\mbox{\boldmath$\sigma$} \psi | \eta_c\rangle $ is suppressed by $v^{7/2}$. 
This suppression factor can be deduced using the methods described in
Ref.\cite{Braaten}. In Coulomb gauge, the leading contribution to this matrix
element comes from a $|c\bar{c}g\rangle $ Fock state, with the $c\bar{c}$ in a
color-octet $^3S_1$ state. This Fock state is dominated by dynamical gluons
with momenta of order $m v$ and has a probability of order $v^3$. Relative to
$\langle 0 | \chi^\dagger \psi | \eta_c \rangle $, the matrix element 
$\langle 0 | \chi^\dagger {\bf B}\cdot \mbox{\boldmath $\sigma$} 
\psi | \eta_c \rangle $ scales like $(mv)^2 v^{3/2}$, with the factor of
$v^{3/2}$ coming from the amplitude for the $c\bar{c}g$ Fock state and the
factor of $(mv)^2$ coming from the dimension of the operator ${\bf B}$. Thus
the suppression factor is $v^{7/2}$.
Keeping only the terms through 
relative order $v^2$, the $T$-matrix element reduces to 
\begin{eqnarray} 
 {\cal T}_{\eta_c \to \gamma\gamma} &\;=\;& - i  \;
 {4  Q^2  e^2  M_{\eta_c} \over  M_{\eta_c}^2 + 4m^2   } \; 
 \mbox{\boldmath $\epsilon$}_1 \times \mbox{\boldmath $\epsilon$}_2 
 \cdot {\bf \hat{r} }\,  
  \nonumber \\
   && \;
   \times \; \Bigg\{ \; 
   \langle 0 | \chi^\dagger  \psi | \eta_c \rangle
     \; + \;
     {3 M_{\eta_c}^4 + 16  m^2 M_{\eta_c}^2 + 144 m^4 
      \over  6m^2(M_{\eta_c}^2 + 4 m^2)^2 } \;  
   \langle 0 | \chi^\dagger {\bf D}^2  \psi |\eta_c \rangle  
 \; \Bigg\} \;.
\label{expan-A-etac:4}
\end{eqnarray}

We obtain the decay rate by squaring the  $T$-matrix element, summing over 
the polarizations of the photons,  and integrating over their phase space:
\begin{eqnarray}
{\Gamma}(\eta_c \to \gamma\gamma) &\;=\;&  
 {16 \pi Q^4  \alpha^2 \; M_{\eta_c}  \over ( M_{\eta_c}^2 + 4 m^2  )^2 } \; 
   \nonumber \\
   && \;
   \times \, \Bigg| \; 
   \langle 0 | \chi^\dagger  \psi | \eta_c \rangle
 \;+\; 
      {3 M_{\eta_c}^4 + 16  m^2 M_{\eta_c}^2 + 144 m^4 
      \over  6m^2(M_{\eta_c}^2 + 4 m^2)^2 } \;  
   \langle 0 | \chi^\dagger {\bf D}^2  \psi |\eta_c \rangle  
 \; \Bigg|^2 \;.
\label{Wid-etagg:3}
\end{eqnarray}
The factorization formula (\ref{Wid-etagg:3}) can be greatly simplified by
using the identity analogous to (\ref{Mpsi-m}) for the $\eta_c$:
\begin{equation}
{M_{\eta_c} \over 2 m} \;=\; 1 \,-\, { 1\over 2 m^2}\,  
 {  \langle 0 | \chi^\dagger {\bf D}^2  \psi | \eta_c \rangle  
  \over
   \langle 0 | \chi^\dagger \psi | \eta_c \rangle } \,
   +\,  O(v^4)  \;.
\label{Metac-m}
\end{equation}
Eliminating the quark mass from the coefficient of the leading matrix 
element in (\ref{Wid-etagg:3}) and keeping only terms through relative order
$v^2$, we get 
\begin{eqnarray}
{\Gamma}(\eta_c \to \gamma\gamma) &\;=\;&  
 {4 \pi Q^4  \alpha^2   \over  M_{\eta_c}^3 } \; 
% \nonumber \\  && \; \times \, 
   \Bigg| \; 
   \langle 0 | \chi^\dagger  \psi | \eta_c \rangle
 \;+\; 
      {1 \over 6 m^2 } \,  
   \langle 0 | \chi^\dagger {\bf D}^2  \psi |\eta_c \rangle  
 \; \Bigg|^2 \;.
\label{Wid-etagg:3p}
\end{eqnarray}
The matrix elements can be expressed in terms of those introduced in 
Ref.\cite{B-B-L} by using (\ref{norm:bbl}). 
Using (\ref{Metac-m}) to eliminate the remaining factors of $M_{\eta_c}$ in favor of the
quark mass,  we recover
the factorization formula derived in Ref.\cite{B-B-L} up to corrections that are
higher order in $v$:
\begin{eqnarray}
{\Gamma}(\eta_c \to \gamma\gamma) &\;=\;&  
 {2 \pi Q^4  \alpha^2   \over  m^2 } \; 
% \nonumber \\  && \; \times \, 
   \Bigg| \; 
   \langle 0 | \chi^\dagger  \psi | \eta_c \rangle_{\rm\small BBL} 
 \;+\; 
      {2 \over 3 m^2 } \,  
   \langle 0 | \chi^\dagger {\bf D}^2  \psi |\eta_c \rangle_{\rm\small BBL}  
 \; \Bigg|^2 \;.
\label{Wid-etagg:4}
\end{eqnarray}
%

%%%%%%%%%%%%%%%%%  
\subsection{ Quarkonium-to-Vacuum Matrix Elements  }
In order to deduce the renormalon ambiguities in the  NRQCD matrix elements
in the factorization formulas (\ref{Wid-psiee:3}) and (\ref{Wid-etagg:3p}), 
it will prove useful to also consider quarkonium-to-vacuum matrix elements 
for all the QCD operators with dimension 3. A complete set of these operators 
is the scalar density $\bar{\Psi}\Psi$, the vector current 
$\bar{\Psi} \gamma^\mu \Psi$, the tensor current $\bar{\Psi} \sigma^{\mu \nu} \Psi$,
the  axial vector current $\bar{\Psi} \gamma^{\mu} \gamma_5 \Psi$, 
and the  pseudoscalar density $\bar{\Psi} \gamma_5 \Psi$. 
Using the FWT transformation (\ref{2-4}), we expand the  matrix elements
of each of these operators in terms of NRQCD matrix elements. 
Keeping terms up to second order in ${\bf D}$, 
the matrix elements for a general quarkonium state $H$ are 
\begin{eqnarray}
 \langle 0 | \bar{\Psi}  \Psi | H \rangle
  &\;=\;& {i \over m } \;
   \langle 0 | \chi^\dagger {\bf D}\cdot \mbox{\boldmath$\sigma$} 
   \psi | H \rangle \;, 
  \\ 
 \langle 0 | \bar{\Psi} \gamma^\mu  \Psi | H \rangle
  &\;=\;&  \delta^{\mu i} \;
   \langle 0 | \chi^\dagger \,\left[\, \sigma^i  
  \;+\; {1 \over 4m^2}  \,
    (D_i D_j \,+\, D_j D_i )\,\sigma^j \,
  \,\right]\,\psi | H \rangle  \;, 
 \\
  \langle 0 | \bar{\Psi} \gamma^{\mu}\gamma_5 \Psi | H \rangle 
 &\;=\;& \delta^{\mu 0} \, 
  \langle 0 | \chi^\dagger \,\left[\, 1 
  \;+\; {1 \over 2m^2}  \,(\,{\bf D}^2  
   + g_s {\rm\bf B} \cdot \mbox{\boldmath $\sigma$} \,)\,  
    \right]\,\psi |H \rangle    
 \nonumber \\ && \hspace{-0.3in} 
 \;+\; \delta^{\mu i}\, { 1\over m}\, \langle 0 | \chi^\dagger \, 
      \left( {\rm\bf D} \times \mbox{\boldmath $\sigma$} \right)^i 
    \psi | H \rangle  \;,
  \\
 \langle 0 | \bar{\Psi} \sigma^{\mu \nu} \Psi |  H \rangle
 &\;=\;&  \delta^{\mu i} \delta^{\nu j}\, {i \over m } \,\epsilon^{ijk} \, 
  \langle 0 | \chi^\dagger D_k \psi | H \rangle
\;-\; i (\delta^{\mu 0} \,\delta^{\nu i} - \delta^{\nu 0} \,\delta^{\mu i})    
 \nonumber \\ &&\hspace{-0.3in} 
    \times\; \langle 0 | \chi^\dagger\,\left[\, \sigma^i 
  \;+\; {1 \over 4 m^2} \, 
    \left(\, 2 {\bf D}^2 \sigma^i \, 
    - (D_i D_j + D_j D_i )\, \sigma^j \, 
     + 2 g_s B^i \,\right) \, 
  \right] \, \psi | H \rangle \;, 
   \\ 
  \langle 0 | \bar{\Psi} \gamma^5 \Psi | H \rangle  
 &\;=\;& - \, \langle 0 | \chi^\dagger  \psi | H \rangle
 \; .
\end{eqnarray}

We now specialize to the $J/\psi$ and the $\eta_c$. 
For the $J/\psi$, the only operators  
that have  matrix elements of order $v^{3/2}$ are the vector current, whose 
matrix element is given in (\ref{A-expan:2}), and   
the tensor current. Keeping only term through relative order $v^2$, the 
matrix element of the tensor current is 
\begin{eqnarray} 
 \langle 0 | \bar{\Psi} \sigma^{\mu \nu} \Psi | J/\psi \rangle
 &\;=\;&  -i (\delta^{\mu 0} \,\delta^{\nu i} - \delta^{\nu 0} \,\delta^{\mu i}) 
  \;\Bigg[ \;
  \langle 0 | \chi^\dagger \sigma^i \psi |J/\psi \rangle
  \;+\; {1 \over 3m^2}  
  \langle 0 | \chi^\dagger {\bf D}^2 \sigma^i \psi |J/\psi \rangle 
   \;\Bigg] \;.
\label{Tensor-expan}
\end{eqnarray}
For the $\eta_c$, the operators that have matrix elements of order 
$v^{3/2}$ are the axial vector current and the pseudoscalar density. 
Keeping only term through relative order $v^2$,  these matrix elements are 
\begin{eqnarray} 
  \langle 0 | \bar{\Psi} \gamma^{\mu}\gamma_5 \Psi | \eta_c \rangle 
 &\;=\;& \delta^{\mu 0} \;\Bigg[ \; 
  \langle 0 | \chi^\dagger  \psi | \eta_c \rangle
  \;+\; {1 \over 2m^2} 
  \langle 0 | \chi^\dagger {\bf D}^2  \psi |\eta_c \rangle  
\; \Bigg] \;, 
\label{Axial-expan} \\
  \langle 0 | \bar{\Psi} \gamma^5 \Psi | \eta_c \rangle  
 &\;=\;& -  \langle 0 | \chi^\dagger  \psi | \eta_c \rangle
 \; .
\label{Pseudo-expan}
\end{eqnarray}
%

%%%%%%%%%%%%%%%%%%%%%%%%%%%%%%%%%%%%%%%%%%%%%%%%
\section { short-distance coefficients in the large-$N_{\small f}$ limit}  
In this section, we briefly review the Borel resummation method and the
standard prescription for calculating the Borel transform of a perturbation
series in QCD in the large-$N_f$ limit. We apply this prescription to the pole
mass to illustrate the renormalization of the Borel transform. 
We then use  the threshold 
expansion method to calculate the Borel transforms of the short-distance
coefficients of the leading matrix elements in the NRQCD  
factorization formulas for   
$ \langle 0 | \bar{\Psi} \gamma^{\mu} \Psi |J/\psi \rangle $, 
$ \langle 0 | \bar{\Psi} \sigma^{\mu\nu} \Psi |J/\psi \rangle $, 
$ \langle 0 | \bar{\Psi} \gamma^{\mu} \gamma_5 \Psi | \eta_c \rangle $, 
$ \langle 0 | \bar{\Psi} \gamma_5 \Psi | \eta_c  \rangle$, 
and  the decay amplitude for $\eta_c\to \gamma \gamma$. 

%%%%%%%%%%
\subsection{Borel Resummation} 
If a function $f(\alpha_s)$ has the power series expansion 
\begin{equation} 
f(\alpha_s) \;=\;  \sum_{n=0}^{\infty}\, a_n \alpha_s^{n} \;, 
\label{original}
\end{equation} 
its Borel transform is defined by
\begin{equation} 
\widetilde{f}( t ) \;=\; a_0 \delta(t) \;+\; \sum_{n=1}^{\infty} 
 { 1 \over (n-1)! }\;  a_{n} t^{n-1} \;. 
\label{borel} 
\end{equation} 
Obviously, the power series (\ref{borel}) for the Borel 
transform has better convergence properties than that of $f$. 
The original function $f$  can be recovered from 
its Borel transform $\widetilde{f}( t ) $ by  
the inverse Borel transformation:  
\begin{equation} 
f(\alpha_s) \;=\; \int_0^{\infty} \;dt \; e ^{-t/\alpha_s} \; \widetilde{f}
(t)\;. 
\label{borel-inverse} 
\end{equation} 
If the integral (\ref{borel-inverse}) is sufficiently convergent,  it 
defines a unique function with the power series expansion (\ref{original}) 
and  the series is called 
Borel resummable. 

A singularity  in the Borel parameter $t$ is called a renormalon\cite{thooft}. 
The contribution to $f(\alpha_s)$ from 
a pole of $\widetilde{f}( t )$ 
at $t=t^*$ with residue $R^*$ can be
obtained by replacing $\widetilde{f}( t )$ in (\ref{borel-inverse})
by $R^*/(t-t^*)$. Expanding as a power series in $t$ and then integrating, we
get
\begin{equation}
f(\alpha_s) \;\sim\; - R^* \, \sum_n \, (n-1)!\, 
\left(\, {\alpha_s \over t^*}\,\right)^{n} \;. 
\label{expan:renorm} 
\end{equation} 
Thus a renormalon gives rise to $n!$ growth of the
coefficients in the power series for $f(\alpha_s)$. 

If the renormalon is on the negative real axis, this contribution to the
series is Borel resummable. 
If the  renormalon is on the positive real axis,  there is an ambiguity in the
function $f(\alpha_s)$. 
The possible prescriptions for the inverse Borel transform include   
deforming the integration contour in  (\ref{borel-inverse}) into the complex
plane  so that it runs above  the pole or below the pole. We also can 
take a linear combination of these two contours, such as the principle value. 
These prescriptions all differ by amounts that are proportional to the
residue of the integrand of (\ref{borel-inverse}) at the pole. 
 If the pole in $\widetilde{f} (t)$  is at the point $t^*$ and has
residue $R^*$, the ambiguity in $f(\alpha_s)$ has the form 
\begin{equation}
\Delta f (\alpha_s) \;=\; K\, (2 \pi \,R^* ) 
      \, e^{- { t^* / \alpha_s  } } \;,
\label{ambigu}
\end{equation}
where $K$ is an arbitrary constant. 
If $\alpha_s$ is the running coupling constant of QCD, its 
scale dependence is approximately 
\begin{equation}
\alpha_s (Q) \approx 
{1 \over \beta_0 \ln ({Q^2 / \Lambda_{\small\rm QCD} ^2 }) },
\label{alphas}
\end{equation}
where $\beta_0=(33-2N_f)/ 12\pi$. Inserting this into (\ref{ambigu}), we see
that the renormalon ambiguity can be approximated by 
\begin{equation}
\Delta f \;\approx \; K ( 2 \pi \,R^*)\,\left({\Lambda_{\small\rm QCD} ^2  
\over Q^2 }  \right)^{ \beta_0 t^*} \;.
\label{ambigu:2}
\end{equation}
Thus the ambiguity is suppressed by a power of $Q^2$. 

To study the  renormalons in NRQCD factorization formulas, we need  
to calculate the short-distance coefficients to all orders  in $\alpha_s$. 
An exact calculation of the coefficients to all orders would be impossible. 
However the calculation is tractable in the large-$N_f$ limit of QCD, where 
$N_f$ is the number of light quarks\cite{Beneke}. The diagrams that dominate 
in this limit are obtained by inserting a chain of 1-loop quark bubbles into 
the gluon propagator in the lowest order diagrams that contain a gluon. 
All other diagrams  are suppressed by a factor of $1/N_f$.  
QCD is no longer an asymptotically free theory 
when $N_f\to \infty$, so it is  more appropriate to consider 
the limit $N_f \to - \infty$. 
To be explicit, we define the large-$N_f$ limit to be the limit $N_f \to 
-\infty$, $\alpha_s \to 0 $, with $N_f \alpha_s$ fixed. 

The insertion of a light quark bubble renormalized by  minimal subtraction
into the propagator of 
a gluon of momentum $\l$ gives  a multiplicative factor  
\begin{equation}
- \beta_0 \,  \alpha_s  \, 
\left[ {g(\epsilon) \over \epsilon } \, 
\left( {-\l^2 -i\epsilon \over \mu^2 e^{- \Delta} } \right)^{-\epsilon}   
\, - \, {1 \over \epsilon } \right] \;
{ - \l^2 g^{\mu\nu} \,+\, l^\mu l^\nu \over \l^2 + i \epsilon }  \;,
\label{bubble}
\end{equation}
where $\beta_0=-N_f/6\pi$ is the first coefficient of the QCD beta function in
the large-$N_f$ limit,
\begin{equation}
g(\epsilon) \;=\; 
{ 6\,(4\pi )^\epsilon \, \Gamma( 1 + \epsilon)\,  \Gamma^2( 2 - \epsilon)
 \over \Gamma( 4 - 2\epsilon) } \;,
\label{g_epsilon} 
\end{equation} 
and 
\begin{equation}
\Delta = - \gamma + \ln (4\pi) + {5 \over 3} +C \;.
\label{Delta} 
\end{equation} 
The constant $C$  depends on the 
renormalization scheme and has the value $C= -5/3$ in  the 
$\overline{\rm MS}$ scheme. A chain of bubble diagrams forms a geometric 
series. The Borel transform for this series multiplied by $\alpha_s$ 
can be obtained by making the following substitution for the gluon propagator
in covariant gauge:
\begin{eqnarray}
&&\alpha_s\,\left[\, {-g^{\mu\nu} \over \l^2  + i \epsilon} \,+\, 
 (1-\xi)\,{\l^\mu\l^\nu \over (\l^2 + i \epsilon)^2 } \,\right] 
\nonumber 
\\
&& \;\longrightarrow\; 
\exp \left\{ u \left[
	 {g(\epsilon) \over \epsilon } \, 
       \left( {-\l^2 -i\epsilon \over \mu^2 e^{- \Delta} } \right)^{-\epsilon} 
        \,-\, {1 \over \epsilon}
        \right] \right\} 
    {  -\l^2 g ^{\mu\nu} + l^\mu l^\nu  \over (-\l^2 -i\epsilon)^2 }\, 
    -\, \xi \, {\l^\mu\l^\nu \over (-\l^2 - i \epsilon)^2 }
\nonumber 
\\
&& \;=\;
\exp \left\{ u \left[ 
	{g(\epsilon) \over \epsilon} 
      e^{\epsilon (\partial/\partial u')} - {\partial \over\partial u'} 
        - {1 \over \epsilon}
        \right] \right\} \, 
\left.
\left( {\mu ^2 \over e^\Delta} \right)^{u'} 
	{g ^{\mu\nu} \over (-\l^2- i\epsilon)^{1+u'} } 
   \right|_{u'=u} \; 
\;+\; \l^\mu\l^\nu \; {\rm terms},   
\label{gauge}
\end{eqnarray}
where $u=\beta_0 t$ and $t$ is the Borel parameter.  
The diagrams for electromagnetic decays considered in this paper are QED
diagrams up to color factors. 
The contributions from the $l^\mu l^\nu$ terms in (\ref{gauge}) 
cancel by the Ward identities of QED. 
The calculations can therefore be greatly
simplified by  using a propagator whose  Lorentz structure is that of the 
gluon propagator in the Feynman gauge.  
The substitution (\ref{gauge}) then reduces to 
\begin{equation}
\alpha_s \, \,{-g^{\mu\nu} \over \l^2 +i\epsilon } \;\longrightarrow\;
{\cal J}_\epsilon \left( { \mu^2 \over e^C } \right)^u\;
    {g ^{\mu\nu}  \over (-\l^2- i\epsilon)^{1 + u} }  \;,
\label{gauge:FM}
\end{equation}
where the action of the operator ${\cal J}_\epsilon$ 
on a function $f(u)$ is defined by
\begin{eqnarray}
{\cal J}_\epsilon f(u) & \equiv &
\left.
\exp \left\{ u\left[ {g(\epsilon) \over \epsilon} 
e^{\epsilon (\partial /\partial u')} - {\partial \ \over \partial u'}
	- {1 \over \epsilon} \right] \right\} \,
e^{-(\Delta - C)u'} \; f(u') \; \right|_{u'=u} \;. 
\label{J-def} 
\end{eqnarray} 
This operator is defined so that it reduces to the identity operator 
in the limit $\epsilon \to 0$ if it acts on a function with no poles
in $\epsilon$.

In calculating the Borel transform of a sum of bubble chain diagrams, 
the extra factor $1/(-\l^2)^u $ introduced 
in the gluon propagator can introduce divergences in the loop 
integral that do not appear in the one-loop diagram, which corresponds to $u=0$.
Below certain critical values of $u$, new power ultraviolet divergences appear.
At such a value $u^*$, the loop integral has a logarithmic divergence, which
appears as a pole in the Borel transform at $u=u^*$. This pole is an ultraviolet
renormalon. 
Similarly, there are power infrared divergences that appear for $u$
above certain critical values $u^*$. At $u=u^*$, the loop integral is
logarithmically infrared divergent. This logarithmic divergence appears as a
pole in the Borel transform at $u=u^*$, and this is an infrared renormalon.

%%%%%%%%%%%%%%%%%

\subsection {Pole mass and renormalization constant} 
We first calculate the Borel transform of the pole mass and the 
wave function renormalization constant of the heavy quark.  
The self-energy for a heavy quark  with  momentum $p$ can be
expressed as a function of the coupling constant $\alpha_s$,
the matrix $\not \! {p}$, and the bare mass $m_0$. 
We denote the self-energy by $\Sigma(\not \! {p} , \alpha_s)$, 
suppressing the dependence on $m_0$. In perturbation theory, 
the pole mass $m_{\rm pole}(\alpha_s)$ is defined by the pole in the
heavy quark propagator: 
\begin{equation}
 m_{\rm pole}(\alpha_s) \;=\; m_0 \;+\;\Sigma(\not \! {p}, \alpha_s)
 \Bigg|_{\not{p}=m_{\rm pole}(\alpha_s)} \;.
\label{m_p}
\end{equation}
One can solve the equation for $m_{\rm pole}$ order by order in $\alpha_s$.
The on-shell wavefunction renormalization constant $Z(\alpha_s)$ is given by
$Z^{-1} = 1 - \delta Z$, where  
\begin{equation}
\delta Z (\alpha_s) \;=\; 
{ \partial \Sigma(\not \! {p}, \alpha_s ) \over \partial \not \! {p} }
\Bigg|_{\not{p}=m_{\rm pole}(\alpha_s)} \;.
\label{Z-ren}
\end{equation}

A one-loop Feynman diagram for the heavy-quark self-energy  
with a chain of  light quark bubbles inserted is shown in Fig.~3. 
Using the substitution (\ref{gauge:FM}), the Borel transform of 
the sum of the diagrams in Fig.~3 is given by
\begin{equation}
\widetilde{\Sigma} (\not \! {p},t) \;=\; i
{16 \pi \over 3} \; 
{\cal J}_\epsilon
\left( \mu^2 \over e^C \right ) ^{u}
\int_l \; { 1 \over (-\l^2 -i\epsilon )^{1 + u } } 
\;\gamma^\mu \; { 1\over \not \! {p} + \not {\l} - m_0 +i\epsilon } 
\;\gamma_{\mu}  \; ,
\label{E-sef:2} 
\end{equation} 
where the operator ${\cal J}_\epsilon$ is defined in (\ref{J-def}).
To regularize the ultraviolet divergence, we use dimensional regularization 
in $D=4-2\epsilon$ dimensions. The integration measure in (\ref{E-sef:2}) is
\begin{equation}
\int_l \;\equiv\; 
  \left({\mu^2 \over e^\Delta} \right)^\epsilon \,
\int {d ^D l \over { (2\pi) }^D }\;.
\end{equation}
The $\epsilon$-dependent factor in front of the
integral has been inserted so that renormalization by minimal subtraction 
will give the same renormalization scheme parametrized by the constant $C$
used in (\ref{bubble}). The integral
in (\ref{E-sef:2}) can be reduced to an integral over a Feynman parameter. 
For the  Borel transforms of the pole mass in (\ref{m_p}) and 
the wavefunction renormalization constant in (\ref{Z-ren}), 
the integral can be expressed in terms of gamma functions:  
\begin{eqnarray}
\tilde{m} _{\rm pole}(t) &\;=\;& m_0 
   \,\left[\;
   \delta({t}) \, + \,
   {3-2\epsilon \over 3 \pi} \,
   e^{(\gamma-5/3)\epsilon} 
{\cal J}_\epsilon 
\left( {\mu^2 \over e^C m_0^2} \right)^{u+\epsilon}
   { \,  \Gamma( u + \epsilon) \Gamma(3-2 u - 2 \epsilon) 
      \over   (1-2u-2 \epsilon) \, 
      \Gamma( 3 -u -2 \epsilon) } 
      \right] \;, 
\label{Borel-M} \\
\delta \widetilde{Z}(t) &\;=\;& -\,     
    { 3-2\epsilon \over 3 \pi } \,e^{(\gamma-5/3)\epsilon} 
{\cal J}_\epsilon 
\left( {\mu^2 \over e^C m_0^2} \right)^{u+\epsilon}
   { (1+u) \Gamma(u + \epsilon) \Gamma(3-2u - 2 \epsilon) 
        \over (1-2u - 2 \epsilon) \,\Gamma(3-u - 2 \epsilon)}  \;.
\label{Borel-Z}
\end{eqnarray}

The expressions (\ref{Borel-M}) and (\ref{Borel-Z}) are ultraviolet divergent
and require renormalization. A  renormalization procedure for the 
large-$N_f$ limit has been derived by Palanques-Mestre and Pascual\cite{P-P}
and by Broadhurst\cite{Broadhurst}, and  
applied to the renormalization of the Borel transform  
by Beneke and Braun\cite{BB94}.  
We can simplify their 
discussion  since the renormalization of the light quark bubble diagram has 
already been carried out in (\ref{bubble}). 
The ultraviolet  divergences in (\ref{Borel-M}) 
and (\ref{Borel-Z}) appear as poles in $\epsilon$ in the coefficients 
obtained by expanding these
expressions as power series in $u$. 
The poles in $\epsilon$ arise from the power
series expansion of the factor $\Gamma(u +\epsilon)$. In the minimal
subtraction renormalization scheme, renormalization is accomplished by
subtracting the pole terms from the coefficients and then taking the limit
$\epsilon\to 0$.
The functions of $u$ in (\ref{Borel-M}) and (\ref{Borel-Z})
can be expressed in the form 
\begin{equation} 
\widetilde{f}(u,\epsilon) \;=\; 
{\cal J}_\epsilon {F(u+\epsilon,\epsilon) \over u+\epsilon}, 
\label{f-bare}
\end{equation}
where $F(u+\epsilon,\epsilon)$ is analytic in $u+\epsilon$ at 
$u+\epsilon =0$. 
The renormalization of this expression by minimal subtraction
involves subtracting the poles in $\epsilon$ from the coefficients 
in its power series expansion in $u$, followed by taking the limit 
$\epsilon \to 0$.  The final result is rather simple:
\begin{equation}
\widetilde{f}_{MS}( u) \;=\; 
{1 \over u } \left[ F(u,0) - \widetilde{G}(u) \right] \;,
\label{f-renorm} 
\end{equation}
where $\widetilde{G}(u)$ is the  Borel transform of the function
\begin{equation}
G(x) \;=\; {(4 \pi)^x \Gamma(4+2x) 
		\over 6 \Gamma(1-x) \Gamma^2(2+x) } F(0,-x)\;.
\label{G-def} 
\end{equation}
We can derive this result through several simple steps that allow us 
to take the limit $\epsilon \to 0$ in parts of the expression
(\ref{f-bare}).  The first step is to subtract 
$e^{(\Delta - C)\epsilon} F(0,\epsilon)/(u'+\epsilon)$ from
$e^{-(\Delta - C)u'} F(u'+\epsilon,\epsilon)/(u'+\epsilon)$ 
in (\ref{f-bare}), and add it back in again.  
The subtraction cancels the pole at $u'+ \epsilon = 0$, so
we can take the limit $\epsilon \to 0$ except in the added term:
\begin{eqnarray}
\widetilde{f}( u, \epsilon)  &\longrightarrow&
{ F(u,0) - e^{u(\Delta - C)} F(0,0) \over u }
\nonumber 
\\
&& \;+\; F(0,\epsilon) \, e^{-u /\epsilon + (\Delta - C)\epsilon} \,
\left.
\exp \left\{ u \left[ {g(\epsilon) \over \epsilon  } 
e^{\epsilon (\partial/\partial u') } - {\partial \over\partial u'} 
 \right] \right\} \,
\, {1 \over  u' + \epsilon} 
\right|_{u'=u} \;.
\label{S-F}
\end{eqnarray}
It is convenient to use the following integral representation for 
$1/(u'+ \epsilon)$:
\begin{equation}
{1 \over  u' + \epsilon}
\;=\;
\int^\infty_0 \, d \, \alpha \, e^{-(u ' + \epsilon ) \alpha }
\end{equation}
Inside the integral, the action of the differential operator 
$\partial/\partial u'$ is simply multiplication by $- \alpha$.
After setting $u'=u$, the action of the exponentiated operator 
in (\ref{S-F}) can be carried out analytically:
\begin{equation}
\left. \exp\left[ u  {g(\epsilon) \over \epsilon  } 
e^{\epsilon (\partial/\partial u') } - u {\partial \over\partial u'} 
\right]  \,
\, {1 \over  u' + \epsilon} 
\right|_{u'=u} 
\;=\; {1 \over u g(\epsilon)}
\left[ e^{ug(\epsilon)/\epsilon} \;-\; 1 \right].
\label{int-exp}
\end{equation}
Inserting (\ref{int-exp}) into (\ref{S-F}),
we can take the limit  $\epsilon \to 0$ in the 
term with the factor $e^{ug(\epsilon)/\epsilon}$.
The resulting expression for the Borel transform is
\begin{equation}
\widetilde{f}( u, \epsilon)  \;\longrightarrow\;
{1 \over u} \left[ F(u,0) \;-\;  
 	{F(0, \epsilon) \over g(\epsilon)}\, 
 	e^{-u/\epsilon + (\Delta - C) \epsilon} \right] \,
\label{f-tilde}
\end{equation}
Minimal subtraction can now be carried out by using the identity
\begin{equation}
\left[ G(-\epsilon) e^{-u/\epsilon} \right]_{MS}  \;=\;
\widetilde{G}(u) \; ,
\label{ren-S}
\end{equation}
where $\widetilde{G}(u)$ is the Borel transform of the function $G(x)$.
Inserting this into (\ref{f-tilde}), we obtain our final result 
(\ref{f-renorm}).

We now apply (\ref{f-renorm}) to the expression (\ref{Borel-M}) 
to obtain an expression for the Borel transform of the pole mass 
in terms of the running mass $m(\mu)$ defined by minimal subtraction:
\begin{equation}
{\tilde m}_{\rm pole}(t) \,=\, m({\mu})  
   \,\Bigg\{\, \delta({t}) 
\,+\, { 1 \over  \pi u } \, 
   \Bigg[\, \left({ \mu^2\over e^C m^2(\mu) } \right)^{u}  \, 
   \, {  \Gamma(1+ u ) \Gamma(3-2 u ) \over (1-2u ) \, \Gamma( 3 -u ) }\,
    \,-\, \widetilde{G}_m(u)  \,
    \Bigg]\,
    \Bigg\} \;,
\label{ren-M} 
\end{equation}
where $\widetilde G_m(u)$ is the Borel transform of the function 
\begin{equation}  
  G_m(x) \;= \; {(3+2x) \Gamma( 4 +2x) 
  \over 9 \Gamma( 1 -x)\,  \Gamma^2( 2 +x) \Gamma(3+x)} \;.
\label{Gm}
\end{equation}
This agrees with Eqs. (4.3) and (4.4) of Ref.\cite{BBB}.
The expression (\ref{ren-M}) for ${\tilde m} _{\rm pole} $ has poles in the
Borel parameter $t=u/\beta_0$. These poles are called renormalons.  They
are classified as ultraviolet renormalons or infrared renormalons 
according to whether they  arise from the  integration region  
of high loop momentum  or low loop momentum. 
The pole in (\ref{ren-M}) from high loop momentum are those that appear in the
factor $\Gamma(1+u)$. 
Thus there are ultraviolet renormalons in the pole mass at the negative 
integers $u=-1,-2,\cdots$. There are infrared renormalons 
at the positive half-integers  $u = {1\over 2}$, 
$ {3\over 2}$, $\cdots$, and also at $u=2$. 
	From (\ref{Borel-Z}), we see that  $\delta Z$ 
has renormalons at the same positions as the pole mass 
except that there is no renormalon at $u=-1$. 

%%%%%%%%%%%%%%%%%%
\subsection {Quarkonium-to-Vacuum Matrix Elements}
Now we calculate the Borel transforms of
the short-distance coefficients in the NRQCD 
factorization formulas for the quarkonium-to-vacuum matrix elements.
We define the short-distance coefficients $C^\Gamma(\alpha_s)$,
where $\Gamma = V,\, A,\, T,\, P $, by the factorization formulas
\begin{eqnarray}
 \langle 0 | \bar{\Psi} \gamma^{\mu} \Psi |J/\psi \rangle 
 &\;=\;& \delta^{\mu i} \,\left[\, 1+C^V(\alpha_s) \,\right] \,
 \langle 0 | \chi^\dagger \sigma^i \psi | J/\psi \rangle \;+\; \cdots \;, 
\label{Coe-V}
\\
 \langle 0 | \bar{\Psi} \gamma^{\mu} \gamma_5 \Psi | \eta_c \rangle  
 &\;=\;& \delta^{\mu 0} \, \,\left[\, 1+C^A(\alpha_s) \,\right] \,
 \langle 0 | \chi^\dagger  \psi | \eta_c \rangle \;+\; \cdots \;, 
\label{Coe-A}
\\
 \langle 0 | \bar{\Psi} \sigma^{\mu\nu} \Psi |J/\psi \rangle 
 &\;=\;&  - i (\delta^{\mu 0} \,\delta^{\nu i} 
 	- \delta^{\nu 0} \,\delta^{\mu i}) \,
\,\left[\, 1+C^T(\alpha_s) \,\right] \, 
 \langle 0 | \chi^\dagger \sigma^i \psi | J/\psi \rangle \;+\; \cdots \;,
\label{Coe-T}
\\ 
 \langle 0 | \bar{\Psi} \gamma_5 \Psi | \eta_c  \rangle 
&\;=\;& - \,\left[\, 1+C^P(\alpha_s) \,\right] \,\langle 0 | \chi^\dagger \psi | \eta_c \rangle \;+\; \cdots \;.
\label{Coe-P}
\end{eqnarray} 
We use $\Gamma$ to represent the matrices $\gamma^\mu$, 
$\gamma^\mu\gamma_5$, $\sigma^{\mu\nu}$, and $\gamma_5$, 
so that the matrix elements on the left 
side of (\ref{Coe-V})-(\ref{Coe-P}) are denoted genericly by 
$ \langle 0 | \bar{\Psi} \Gamma \Psi | H \rangle $. 

The short-distance coefficients $C^\Gamma(\alpha_s)$ in
(\ref{Coe-V})-(\ref{Coe-P}) can be determined by matching the 
corresponding equations for $|c\bar{c} \rangle$ states computed in
perturbation theory. The simplest choice is a state $|c\bar{c}\rangle$ 
that consist of a $c$ and $\bar{c}$ at rest. We must compute  
$\langle 0|\bar{\Psi} \Gamma \Psi | c\bar{c} \rangle $ using perturbative 
QCD and $\langle 0| \chi^\dagger \sigma^i  \psi | c \bar{c} \rangle $ and 
$\langle 0| \chi^\dagger  \psi  | c \bar{c} \rangle $
using perturbative NRQCD, and then determine  $C^\Gamma(\alpha_s)$
by matching the left and right sides of 
(\ref{Coe-V})-(\ref{Coe-P}). By Borel
transforming both sides of the equation before carrying out the matching, we
can determine the Borel transform ${\widetilde C}^\Gamma(t)$.   

We first consider the QCD side of the matching conditions. 
The $c$ and $\bar{c}$ at rest have equal four-momenta $p=(m,{\bf 0})$ 
and are represented by the spinors 
\begin{eqnarray}
&&
u(p)\;=\; \sqrt{2m}\,
{\xi \choose 0} \,
,~~~~~ 
v(p)\;=\; \sqrt{2m}\,
{0 \choose  \eta}
 \;.
 \label{spinors}
\end{eqnarray}
The diagrams that dominate the radiative corrections in the large-$N_f$ 
limit are obtained by inserting a chain of 
light quark bubbles into the gluon propagator in the one-loop diagrams. 
The diagrams with propagator corrections on the $c$ and $\bar{c}$ 
lines reduce  to the tree level matrix element 
$\bar{v}\Gamma u$ multiplied by $\delta Z(\alpha_s)$. After Borel 
transforming, they give $\delta\widetilde{Z}(t) \bar{v}\Gamma u$. 
The remaining diagrams are vertex corrections, such as the diagram for
$c\bar{c} \to e^+ e^-$ in Figure 4.  The Borel transform 
of the vertex correction is
\begin{eqnarray}
&&  
 i\,{16\pi \over 3} \;    
{\cal J}_\epsilon \left( {\mu^2 \over e^C} \right ) ^{u} 
\int_l \;{ 1 \over (-\l^2 - i \epsilon )^{1 + u} } \;
\bar{v}(p)  \,
\;\gamma^\nu \;   
{ 1\over \not {\l} - \not \! p - m + i \epsilon} \;
\Gamma \; 
{ 1\over \not {\l} + \not \! p - m + i \epsilon} \;
\gamma_\nu  \,
u(p) \;.
\label{Borel-G:2} 
\end{eqnarray}
The dimensionally regularized integral in (\ref{Borel-G:2}) can be 
evaluated by introducing Feynman parameters. After  a straightforward 
calculation, we find that (\ref{Borel-G:2}) reduces to  
${\tilde D}^{\Gamma} (t) \, \bar{v}(p)\Gamma u(p)$, where  
\begin{eqnarray}
{\tilde D}^{\Gamma} (t) \,
&\;=\;& {2 \over 3 \pi}   
 e^{(\gamma-5/3)\epsilon} \;
{\cal J}_\epsilon \left( {\mu^2 \over e^C m^2 } \right)^{u+\epsilon}\; 
   { \Gamma(u + \epsilon) \Gamma(1-2u - 2 \epsilon) 
         \over (1+2u + 2 \epsilon) \, 
         \Gamma(3-u -2 \epsilon) } 
\nonumber
\\
& &  %\hspace{-1.2in}
  \times \;
   \left\{ \;\;\;
    \begin{array}{ll} 
      (1-2\epsilon - u) \,[(1-\epsilon)(3+2\epsilon) \,+\, (1-2\epsilon)u \,]
        & ~~~~~~~~\Gamma=V  \;,\label{D-V} \\
      (3-2\epsilon) \, [(1-\epsilon) \,+\, \epsilon u + u^2 \,]
       & ~~~~~~~~\Gamma=A  \;,\label{D-A} \\ 
      2(1-\epsilon)(1-2\epsilon) \,-\, (3-3\epsilon +2 \epsilon^2) u 
      \,+\,(1-2\epsilon) u^2  
       & ~~~~~~~~\Gamma=T  \;,\label{D-T} \\ 
      (3-2\epsilon)(1+\epsilon+u)(2-2\epsilon  - u)  
       & ~~~~~~~~\Gamma=P  \;, \label{D-P} 
     \end{array} 
    \right. 
\label{Borel-G:3} 
\end{eqnarray}
where the operator ${\cal J}_\epsilon$ acts on every function of
$u$  to the right.
The prescription we have used for $\gamma_5$ in $D$ dimensions is that 
$\{ \gamma_5, \gamma^\mu \} =0 $ for all $\mu$.

We next consider the NRQCD sides of the matching conditions. If we use
dimensional regularization, the only scale in the perturbative NRQCD 
matrix elements is the relative
momentum ${\bf q}$. Having set ${\bf q}=0$, the only matrix elements that
survive are  $\langle 0 | \chi^\dagger \sigma^i  \psi | c\bar{c} \rangle $
and $\langle 0 | \chi^\dagger  \psi | c\bar{c} \rangle $.
The matrix elements of all higher dimension operators vanish. The 
radiative corrections to the  matrix elements vanish in dimensional
regularization, since there is no momentum scale in the loop integral. Thus the
NRQCD matrix elements are trivial:
\begin{eqnarray}
\langle 0 | \chi^\dagger \sigma^i  \psi | c\bar{c} \rangle
&\;=\;&
 {2m}\, \eta^\dagger \sigma^i \xi \;,
 \label{ccbar-vector}
\\
\langle 0 | \chi^\dagger  \psi | c\bar{c} \rangle
&\;=\;&
{2m}\, \eta^\dagger  \xi \;.
\label{ccbar-pseudo}
\end{eqnarray}
Since these matrix elements are independent of $\alpha_s$, the Borel 
transforms of the right sides of (\ref{Coe-V})-(\ref{Coe-P}) are 
obtained simply by replacing $ 1 +C^\Gamma(\alpha_s)$ by 
$\delta(t) + \widetilde{C}^\Gamma(t)$.

By matching the Borel transforms of the QCD and NRQCD sides of
(\ref{Coe-V})-(\ref{Coe-P}), we find that the short-distance 
coefficients are given by 
\begin{equation}
\widetilde{C} ^{\Gamma} (t) \;=\;
{\tilde D} ^{\Gamma} (t)  \;+\; \delta \widetilde{Z}(t)
\;,
\end{equation}
where $ \delta\widetilde{Z}(t) $ is given in (\ref{Borel-Z}) and 
${\tilde D} ^{\Gamma} (t)$ for $\Gamma = V,\, A,\, T$, and 
$P$ is given in (\ref{Borel-G:3}). 
The Borel transforms $ \widetilde{C}^{V } $ and 
$\widetilde{C}^{A} $ are finite at $u+\epsilon=0$  
and therefore do not need to be renormalized.  
This is simply a consequence of the conservation of the vector
current and  the partial conservation of the axial current. 
We can therefore set $\epsilon =0 $ in these coefficients: 
\begin{eqnarray} 
\widetilde{C}^{V}( t)  &\;=\;& 
    -\,{2 \over 3 \pi} \; 
    \left( {\mu^2 \over e^C m^2} \right ) ^{u } \;
      { 4+3u \over (1-2u)(1 +2u) } \;
   { \Gamma( 1+ u ) \, \Gamma(3-2 u ) \over  \Gamma( 3 -u  ) } \;,
    \label{C-V}\\
\widetilde{C}^{A}( t)  &\;=\;& 
    -\,{4 \over  \pi} \; 
    \left( {\mu^2 \over e^C m^2} \right ) ^{u } \;
      {  1-u-u^2  \over 1+2u } \;
   { \Gamma( 1+ u ) \, \Gamma(1-2 u ) \over \Gamma( 3 -u  ) } \;.
    \label{C-A}
\end{eqnarray}
In the Borel transforms $\widetilde{C}^T $ and $\widetilde{C}^P $,
there is a pole at $u+\epsilon=0$. This pole indicates that there are
ultraviolet divergences in the coefficients of the power series expansion 
in $u$. These divergences are removed by multiplicative renormalizations 
of the QCD operators $\bar{\Psi} \sigma ^{\mu\nu} \Psi$ and  
$\bar{\Psi} \gamma_5\Psi$.  If we use the minimal subtraction renormalization
scheme, the renormalization can be carried out by making the substitution
(\ref{f-renorm}) [[ CHANGE EQ. NO. (\ref{f-renorm}). ]]
Our final results for these coefficients are 
\begin{eqnarray} 
\widetilde{C}^{T}(t) &=& 
    -\,{1 \over 3 \pi u} \; 
\left[ \left( {\mu^2 \over e^C m^2} \right)^{u} \;
       {1 + 10 u + 6 u^2 \over (1-2u)(1+2u)} \;
   {\Gamma(1+u) \, \Gamma(3-2u) \over \Gamma(3 - u)}
\;-\;  \widetilde G^T(u) \right] \;,
\label{C-T} 
\\
\widetilde{C}^{P}(t)  &=& 
{2 \over \pi u} \; 
\left[ \left( {\mu^2 \over e^C m^2} \right)^{u} \;
       {1-2u+2u^2 \over 1+2u} \;
   {\Gamma(2+u) \, \Gamma(1-2u) \over \Gamma(3-u)}
\;-\;  {1 \over 2} \widetilde G_m(u)  \right] \;,
\label{C-P} 
\end{eqnarray}
where $\widetilde G^T(u)$ is the Borel transform of the function 
\begin{equation}  
  G^T(x) \;=\; {(1+2x) \Gamma( 4 +2x) 
  \over 3 \Gamma( 1 -x)\,  \Gamma^2( 2 +x) \Gamma(3+x)} 
\end{equation}
and $\widetilde G_m(u)$ is the Borel transform of the function 
$G_m(x)$ defined in (\ref{Gm}).

  From (\ref{C-V})-(\ref{C-P}), we see that all the coefficients have  
ultraviolet renormalons at the negative integers 
$u=-1,-2,\cdots$, except for $\widetilde{C}^{P}$, which does not have a
renormalon at $u=-1$. They all have infrared renormalons  at  the positive
half-integers  $u=  {1\over 2}$, $ {3\over 2}$,
$\cdots$, and at $u=2$. The coefficients 
$\widetilde{C}^{A}$ and $\widetilde{C}^{P}$  have an additional 
infrared renormalon at
$u=1$. Finally, all the coefficients have  
a renormalon  at $ u=-{1\over 2 }$. This appears to be
an infrared renormalon in spite of the fact that it appears on 
the negative real axis,  because it  arises from the low momentum region of 
the integral over the loop momentum $\l$ in (\ref{Borel-G:2}). 
However the low momentum region of that integral is actually canceled by the
low momentum region of an integral on the NRQCD side of the matching condition.
The integral on the NRQCD side has the form 
\begin{equation}
-\, {16 \pi m \over 3}\, 
\left( {\mu^2 \over e^\Delta} \right)^\epsilon \,
\int\, {d^{3-2 \epsilon} \l \over (2\pi)^{3-2\epsilon} } \,
{1 \over (\l^2)^{2 +u} } \;. 
\end{equation}
This vanishes in dimensional regularization due to a cancellation between an
infrared pole at $u +\epsilon = -{1 \over 2}$ and an ultraviolet pole at $u
+\epsilon = -{1 \over 2}$. Since the infrared pole is canceled by the QCD
contribution to $\widetilde{C}^\Gamma$, the renormalon at $u=-{1\over 2}$ is
actually an ultraviolet renormalon of NRQCD. 

To obtain the power series expansions for the coefficients
$C^\Gamma(\alpha_s)$, we expand the coefficients $\widetilde{C}^{\Gamma}(t)$
as a power series in $t$ and then invert the Borel transform by substituting 
$t^{n-1} \to (n-1) ! \alpha_s^n $. The first two terms in the expansion of 
$C^V (\alpha_s)$ are 
\begin{equation}
C^V(\alpha_s ) \;=\;
- {8 \over 3 } \, {\alpha_s(\mu) \over \pi } \,+\, 
\left( \,2 - {16 \over 3} \log {\mu \over e^{C/2} m } \,\right)
\, { \beta_0 \alpha_s^2 \over \pi } 
\;.
\label{C-V-2}
\end{equation}
The coefficient of $\alpha_s$ was first calculated by Barbieri 
{ et al}.\cite{wc}. From the $\alpha_s^2$ term in  (\ref{C-V-2}),  
we can read off the Brodsky-Lepage-Mackenzie (BLM) scale\cite{B-L-M} for this process, which is the scale
$\mu$ for which the correction proportional to $N_f\alpha_s^2$
vanishes. We find $\mu= 1.46 \,e^{ C/2} m$.

%%%%%%%%%%%%%%%%%%%% 
\subsection { Decay of $\eta_c $ into $\gamma\gamma$ } 

We next calculate the Borel transform of the short-distance coefficient of
$\langle 0 | \chi^\dagger \psi | \eta_c \rangle$ in the NRQCD factorization 
formula for the $T$-matrix element for $\eta_c \to \gamma\gamma$. 
The factorization formula with tree-level coefficients is given in
(\ref{expan-A-etac:4}). Up to this point, it has not been necessary to define
carefully the mass $m$ that appears in the factorization formula. It could
equally well be the running mass $m(\mu)$ defined by a minimal subtraction
renormalization scheme, or the pole mass $m_{\rm\small pole}$ defined by the
location of the pole in the propagator in perturbation theory, or any other mass
that differs from these by perturbative corrections. However, in order to
calculate the short-distance coefficient beyond tree level, it is necessary to
specify the definition of $m$. The choice is somewhat arbitrary, since a
different choice can be compensated by changes in the perturbative corrections
to the short-distance coefficient. We choose to use the pole mass. 
We define the short-distance coefficient $C^{\gamma\gamma} (\alpha_s)$ 
by the factorization formula
\begin{eqnarray}         
{\cal T}_{\eta_c \to \gamma\gamma} &\;=\;& - i\,
 {4  Q^2 e^2 M_{\eta_c} \over M_{\eta_c}^2 + 4 m^2_{\rm\small pole} } \; 
 \mbox{\boldmath$\epsilon$}_1 \times \mbox{\boldmath$\epsilon$}_2 \cdot
 \hat{\bf r}  
 \; \left[\, 1\,+\,C^{\gamma\gamma}(\alpha_s) \,\right]\; 
\langle 0 | \chi^\dagger \psi | \eta_c \rangle \;+\; \cdots \;, 
\label{C^gg:0} 
\end{eqnarray} 
Our choice of the pole mass is dictated primarily by the algebraic simplicity of
the final result for the Borel transform of $C^{\gamma\gamma}(\alpha_s)$.  
The coefficient $C^{\gamma\gamma}$ in (\ref{C^gg:0}) can be determined 
by matching the perturbative  $T$-matrix elements 
for  $c\bar{c} \to \gamma\gamma$ in QCD and in NRQCD, as 
in (\ref{match}).  If we take the $c$ and $\bar{c}$ to be at rest, their
invariant mass is $M_{c\bar{c}} = 2 m_{\rm pole}$, and the matching condition
reduces to 
\begin{eqnarray}         
{\cal T}_{c\bar{c}\to \gamma\gamma} \Bigg|_{pQCD} &\;=\;&  
-i  {  Q^2 e^2\over 2 m_{\rm\small pole} } \;  
\mbox{\boldmath$\epsilon$}_1 \times \mbox{\boldmath$\epsilon$}_2 \cdot
 \hat{\bf r}  
\left[\,1\,+\, C^{\gamma\gamma}(\alpha_s)\,
 \right]\; 
\langle 0 | \chi^\dagger \psi | c\bar{c} \rangle\Bigg|_{pNRQCD} \;+\; \cdots \;. 
\label{C^gg} 
\end{eqnarray} 
By Borel transforming both sides of the equation, we can determine 
$\widetilde{C}^{\gamma\gamma}(t)$. 

We first consider the NRQCD side of
the matching condition. Having chosen the relative momentum of the $c\bar{c}$
pair to be $0$, the radiative corrections to the matrix element vanish in
dimensional regularization. The matrix element therefore reduces to  
\begin{equation}
\langle 0 | \chi^\dagger \psi | c \bar{c} \rangle = 2 m _{\rm pole} \eta^\dagger
\xi \;.
\end{equation}
The Borel transform of the right side of 
(\ref{C^gg}) is therefore obtained by replacing $1 + C^{\gamma\gamma}(\alpha_s)$ by
$\delta (t) +  \widetilde{C}^{\gamma\gamma}(t) $.

We next consider the QCD side of the matching equation (\ref{C^gg}). Since the
$c$ and $\bar{c}$ are at rest, they have equal 4-momentum $p=(m,{\bf 0})$ and
their spinors
are given in (\ref{spinors}). The diagrams for the $T$-matrix element in the
large $N_f$ limit are those given in Figs. 5a-e, together with the diagrams
with self-energy corrections on the external $c$ and $\bar{c}$ lines. In each of
the diagrams, the spinor factor can be reduced to $ r_\lambda \bar{v}
\Gamma_{-}^{\mu \lambda \nu} u$, where $r=(k_1-k_2)/2$. The contribution to  
$\widetilde{C}^{\gamma\gamma}(t)$ can be obtained from the Borel transform of 
the diagram by removing the factor 
\begin{equation}
 {Q^2 e^2 \over 2 m^2} \epsilon_{1\mu}\epsilon_{2\nu} r_\lambda 
\bar{v}(p) \Gamma_{-}^{\mu\lambda \nu} u(p) \;=\; -i Q^2e^2 \;
\mbox{\boldmath$\epsilon$}_1 \times \mbox{\boldmath$\epsilon$}_2 \cdot
 \hat{\bf r} \, \eta^\dagger \xi \;.
\label{factor}
\end{equation}

The ultraviolet divergences associated with vertex corrections and wavefunction
renormalizations cancel by the Ward identities of QED. The only renormalization that
is necessary is therefore mass renormalization. It is convenient to use on-shell
renormalization. Explicit regularization of the loop integrals is unnecessary, 
since the Borel parameter itself provides a regulator. The mass counterterm 
and the wavefunction renormalization constant can be obtained from 
(\ref{Borel-M}) and (\ref{Borel-Z}) by setting $\epsilon=0$: 
\begin{eqnarray}
\delta \widetilde{m} (t) &\;=\;&  
   m\, {2 \over  \pi }  \;
   \left( { \mu^2\over e^C m^2 } \right)^{ u }  \,
   {1-u \over u } \;
   {   \Gamma( 1- 2 u ) \Gamma(1 + u) 
     \over \Gamma( 3 -u ) } \;, 
 \label{delta-M}
       \\
\delta \widetilde{Z}(t) &\;=\;& -\,     
   {2 \over  \pi }  \;
   \left( { \mu^2\over e^C m^2 } \right)^{ u }  \;
   {(1+u)(1-u) \over u } \;
   {   \Gamma( 1- 2 u ) \Gamma(1 + u) 
      \over \Gamma( 3 -u ) } \;. 
 \label{delta-Z}
\end{eqnarray}

The short-distance coefficient $\widetilde{C}^{\gamma\gamma}(t)$
can be written 
\begin{equation}
\widetilde{C}^{\gamma\gamma}(t) \;=\; 
\widetilde{C}_{1}(t) + \widetilde{C}_{2}(t) +\widetilde{C}_{3}(t) +
\delta \widetilde{Z}(t) \;,
\end{equation}
where $\widetilde{C}_{1}(t) $ is the contribution from the box diagrams 
in Fig. 5a, $\widetilde{C}_{2}(t) $ is the  contribution from the vertex
corrections in Figs.~5b and 5c, 
$\widetilde{C}_{3}(t) $ is the sum of the contribution from the propagator
correction in Fig. 5d and the mass counterterm in Fig 5e, and 
$\delta \widetilde{Z}(t) $ is the contribution from propagator corrections on
the external lines.

The sum of the Borel transform of the box diagram in Fig. 5a 
and the diagram obtained by interchanging the photon lines is given by 
\begin{eqnarray}
 && 
 {16 \pi Q^2 e^2  \over 3} \; \epsilon_{1\mu} \epsilon_{2\nu} 
  \; \left( { \mu^2\over e^C} \right ) ^{u} \; 
\int_l \;{ 1 \over (-\l^2)^{1 + u  } }  \; \bar{v}(p) \gamma^\alpha \; 
{ 1\over \not{\l} - \not\!p - m } \;
\nonumber \\
& & \hspace{0.3in}
 \times\,\left(\; 
   \gamma^\mu \;
     { 1\over \not{\l} + \not{r} - m } \;\gamma^\nu \; 
\,+\,
   \gamma^\nu \;
      { 1\over \not{\l} - \not{r} - m } \;\gamma^\mu \;
\right)\;
{ 1\over \not{\l} + \not\!p - m }\;
\gamma_\alpha u(p) \;. 
\label{Borel-box:1} 
\end{eqnarray}
The integral in Eq.~(\ref{Borel-box:1}) 
can be carried out using Feynman parameters.  
Removing the  factor (\ref{factor}), we obtain
\begin{eqnarray}
\widetilde{C}_{1} (t)  &\;=\;& { 2 \over 3 \pi} \; 
\left( { \mu^2\over e^C m^2 } \right ) ^{u} \;    \; 
  \Bigg[\;
    {(2-u)\,(1+2u + 3u^2) \over u\, (1+u)\,(1+2u) }
      \,{ \Gamma(1- 2u ) \Gamma( 1 + u) \over \Gamma( 3 -u ) } \;
  \nonumber\\& & \hspace{1in} 
  \; -\;{1\over 1+u} \, { \Gamma(1- u ) \Gamma\left( { 1+ u \over 2} \right) 
              \over \Gamma\left( { 3 - u\over 2} \right) } \;       
         \Bigg] \;.
\label{Borel-box:2} 
\end{eqnarray}
 The Borel transform of the vertex correction in Fig. 5b is obtained  by
 replacing $\gamma^\mu$ by 
\begin{eqnarray}
i {16\pi \over 3} \; \left({ \mu^2\over  e^C}  \right ) ^{u} \; 
\int_l \;{ 1 \over (-\l^2)^{1 + u  } }\; \gamma^\alpha \; 
{ 1\over \not{\l} - \not{r} - m } \;\gamma^\mu \;
{ 1\over \not{\l} + \not\!p - m }\;
\gamma_\alpha   \;. 
\label{Borel-vertex:1} 
\end{eqnarray}
Using Feynman parameters to evaluate the integrals  
for the four vertex correction diagrams,
we find that their contribution to
$\widetilde{C}^{\gamma\gamma}$ is 
\begin{eqnarray}
\widetilde{C}_{2} (t)  &\;=\;& {2 \over 3 \pi} \; 
 \left({ \mu^2\over e^C m^2 }  \right ) ^{u} \; \Bigg[\;
  { 2(1-2u) \over u^2 }\;
  { \Gamma( 1 - 2u ) \Gamma(1+ u) \over \Gamma( 3 -u ) } 
  \nonumber\\& & \; 
   \;+\; {1\over u}\, 
   { \Gamma(1- u ) \Gamma\left( { 1+ u \over 2} \right) 
              \over \Gamma\left( { 3 - u\over 2} \right) }
   -{1\over u^2} \,
   { \Gamma(2- u ) \Gamma \left( { 2+ u \over 2} \right) 
                  \over 2\,\Gamma \left( {4 - u\over 2} \right) } 
   \;  \Bigg] \;. 
\label{Borel-vertex:2} 
\end{eqnarray}
The Borel transform  of the self-energy correction for  
the internal quark line  in Fig.~5d is given by (\ref{E-sef:2})  
with $p$ replaced by $-r$. 
With a straightforward calculation, we obtain  
\begin{eqnarray}
\widetilde{\Sigma}_{f} (-\not \!r,  t )  &\;=\;& 
 - {1\over 3 \pi} \;
 \left({ \mu^2\over e^C m^2 }  \right ) ^{u} \; 
  \Bigg[\;
     \not{r} \;
     { \Gamma( - u ) \Gamma \left( {2 +  u \over 2} \right) 
                  \over \Gamma \left( {4 - u\over 2} \right) } 
   \;+\;2m\; { \Gamma(- u ) \Gamma\left( { 1+ u \over 2} \right) 
              \over \Gamma\left( { 3 - u\over 2} \right) } \;     
  \Bigg] \;.
\label{Borel-self:2} 
\end{eqnarray}
The diagram in Fig. 5e has  the mass counterterm  
$\delta \tilde {m}$ inserted into  the internal quark line. 
Adding the contributions to $\widetilde{C}^{\gamma\gamma}$ from  
Figs.~5d and 5e, we obtain 
\begin{eqnarray}
\widetilde{C}_{3} (t)  &\;=\;&  {2 \over 3 \pi} \; 
 \left({ \mu^2\over e^C m^2 }  \right ) ^{u} \; 
  \Bigg[\; -{1\over u}\, 
    { 2 \Gamma(1 - u ) \Gamma\left( { 1+ u \over 2} \right) 
              \over \Gamma\left( { 3 - u\over 2} \right) } \;     
   +\; {3(1-u) \over u} \, {  \Gamma\left( { 1 -2u } \right) \, \Gamma(1+u) 
        \over  \Gamma\left( { 3 - u } \right) } \;     
   \Bigg] \;.
\label{Borel-self:3} 
\end{eqnarray}
Adding up 
(\ref{Borel-box:2}), (\ref{Borel-vertex:2}), (\ref{Borel-self:3}), and
 (\ref{delta-Z}),   the Borel transform of the short-distance coefficient 
for the process $\eta_c \to \gamma \gamma$ is 
\begin{eqnarray}  
 \widetilde{C}^{\gamma\gamma} (t)  & \;=\; & 
 {2\over 3 \pi} \; 
 \left({ \mu^2\over e^C m^2 }  \right ) ^{u} \;  
  \Bigg[\;
      {2(1+2u-4u^2-5u^3+3u^5) \over u^2(1+u)(1+2u) }   \,
     { \Gamma( 1 - 2u ) \Gamma(1+ u) \over \Gamma( 3 -u ) }
  \nonumber\\&  & \hspace{1in} 
    \;-\; {1\over 1+u}\,  { \Gamma(1- u ) \Gamma \left({ 1+ u\over 2} \right) 
           \over \Gamma \left( { 3- u\over 2} \right) } 
   \;-\; {1\over u^2 }\, { \Gamma(1- u ) \Gamma\left( { 2 + u \over 2} \right) 
              \over  \Gamma\left(  {4-  u\over 2} \right) } \;     
  \Bigg] \;. 
\label{Borel-etacgg} 
\end{eqnarray}
While individual terms in (\ref{Borel-etacgg}) have a double pole at $u=0$, 
there are cancellations that make the expression analytic at $u=0$.
	From eq.~(\ref{Borel-etacgg}), we see 
that there are ultraviolet renormalons at the negative integers 
$u=-1,-2,\cdots$. There are  infrared renormalons   
at the even and odd positive half-integers  $u= {1\over 2}$, $1$,  
$ {3\over 2}$, $2$, $\cdots$. Finally,  there is the NRQCD ultraviolet 
renormalon at  $ u=-{1\over 2 }$.

To obtain the power series expansion of $C^{\gamma\gamma} (\alpha_s)$,
we expand (\ref{Borel-etacgg}) in powers of $u$ and carry out the 
inverse Borel transform by substituting $t^{n-1} \to (n-1)! \, \alpha_s^n $. 
The first two terms in the expansion are 
\begin{equation}
C^{\gamma\gamma} (\alpha_s) \;=\;
  { \pi^2 -20 \over 6  } \, 
{\alpha_s(\mu) \over \pi } \,+\, \left( \,5.073 + 
{  { \pi^2 -20 \over 3} \, } \, 
\log {\mu \over e^{C/2} m } \,\right) \, { \beta_0 \alpha_s^2 \over \pi } 
\;.
\label{C-gg-2}
\end{equation}
The coefficient of $\alpha_s$ was first calculated by Barbieri 
{ et al.}\cite{barba}.   From the vanishing of the coefficient of
$\beta_0 \alpha_s^2$, we identify the BLM scale for this process to be  
$\mu= 4.49\,e^{C/2} m$.

%%%%%%%%%%%%%%%%%%%%%%%%%%%%%%%%%%%%%%%%%%%%%%%%%%
\section { Renormalon singularities in quarkonium annihilation decays } 

We have calculated the Borel transforms of the short-distance 
coefficients $C^\Gamma(\alpha_s)$ for $\Gamma=V,A,T,P$, and $\gamma\gamma$
in the large-$N_f$ limit. They have  
renormalon singularities in the form of poles on the real  
axis of the Borel parameter $t=u/\beta_0 $. 
There are ultraviolet renormalons at the points $u=-1$, $-2$, $\cdots$, 
infrared renormalons at ${1\over 2}$, ${3\over 2}$, $2$, ${5\over 2}$, 
$3$, $\cdots$, and a NRQCD ultraviolet renormalon at $u=-{1\over 2}$. 
Each renormalon generates $n!$ growth of the
coefficients in the perturbation series, as shown in (\ref{expan:renorm}). 
The asymptotic behavior of the coefficients is dominated by the renormalons 
closest to the origin at $u=\pm {1\over 2}$. The contributions 
from a higher renormalon at $u={k / 2}$ are suppressed
by $k^{ -n }$. In this section, we study in detail the dominant renormalons at
$u=+{1\over 2}$ and $u=-{1\over 2}$. We show that the Borel-resummable 
renormalon at $u=-{1\over 2}$ arises from the  cancellation of the Coulomb 
singularity in the short-distance coefficients and 
is universal in the sense that its residue is identical for all 
color-singlet $S$-wave decay
processes. We  show that the ambiguities from the renormalon at $u={1\over
2}$ can be absorbed into the nonperturbative NRQCD matrix elements that
contribute through relative order $v^2$.
We also deduce the residues of the renormalons at $u=\pm {1\over 2}$ for the
decay $J/\psi \to \gamma\gamma\gamma$.  

%%%%%%%%%%%%%%%
\subsection  { Universality of the $u=-{1\over 2}$ renormalon } 

The Borel transforms of the short-distance coefficients  
$C^\Gamma(\alpha_s)$ are given in (\ref{C-V})-(\ref{C-P})  for $\Gamma=
V,A,T,P$ and in  (\ref{Borel-etacgg}) for $\Gamma=\gamma\gamma$. 
By examining these expressions, we see that they all have  a 
renormalon at $u=-{1 \over 2}$ and that the residues of the poles 
are identical for all $\Gamma$.  
Thus they can all be written in the form
\begin{equation}
\widetilde{C}^\Gamma (t) \;=\; {1 \over 1 + 2 \beta_0 t }\, 
\left(\, - {8 \over 3 \pi} { e^{ C/2 } \, m \over \mu } \right)\, 
+\, \widetilde{B}^{\Gamma}(t) \;,
\label{C-u:-1/2}
\end{equation}
where $\widetilde{B}^\Gamma (t)$ is analytic at $\beta_0 t= -{1 \over 2}$.
Since the pole at $u = -{1 \over 2}$ is located on the negative real axis, 
the contribution of this renormalon to $C^\Gamma(\alpha_s)$ is 
Borel resummable. If $B^\Gamma (\alpha_s)$ is the inverse Borel  
transform of $\widetilde{B}^\Gamma(t)$, then the
inverse Borel transform of (\ref{C-u:-1/2}) is 
\begin{equation}
C^\Gamma (\alpha_s) \;=\; - {8 \over 3 \pi} \,
{ e^{C/2} \, m \over \mu} \;\alpha_s \, \int^\infty_0 \, 
dx \,{ e^{- x} \over 1 + 2 \beta_0 \alpha_s \, x }
\,+\, {B}^\Gamma (\alpha_s) \;.
\label{C-mu}
\end{equation}
We have suppressed the $\mu$-dependence of the coupling constant
$\alpha_s(\mu)$.  
For sufficiently  small values of $\alpha_s $, the integral is
approximately 1. However its power series expansion in $ \alpha_s $ 
is an alternating series whose coefficients diverge like $n!$.  

We found that the residues of the renormalon at $u = -{1 \over 2}$ 
are the same for all 
processes that we calculated. We will show that this result is general, 
that the residue is identical for all 
color-singlet $S$-wave annihilation decay processes. 
The appearance of this renormalon  can be traced back to a remnant of the 
cancellation of the Coulomb singularity in the short-distance coefficients.   

Any  $S$-wave charmonium decay proceeds through the 
annihilation of the $c\bar{c}$ pair. 
The $c\bar{c}$ annihilation amplitude at tree level 
defines a vertex $\Gamma(p,\bar{p})$, which depends on the 3-momenta of the 
$c$ and $\bar{c}$. The scale of the momentum dependence is $m$. The 
vertex can therefore be approximated by  a momentum independent  vertex  
when $c$ and $\bar{c}$ have 3-momenta that are small compared to $m$. 
The QCD radiative corrections to the annihilation amplitude include the diagram
in Fig.~4  in which  a virtual gluon is exchanged 
between the $c$ and $\bar{c}$. 
If the relative momentum between the $c$ and $\bar{c}$ is zero, 
there is a pinch singularity in the loop integral. 
This is the  Coulomb  singularity. 
If the $c$ and $\bar{c}$ have a small relative momentum $mv$, the singularity
appears as a $1/v$ term in the annihilation amplitudes.  
Since NRQCD by construction matches full QCD at momenta small compared to $m$,
this Coulomb singularity is the same in the two theories. It therefore 
cancels in the short-distance coefficients.
However, the cancellation leaves a finite  
remainder. If we insert a chain of $n$  bubble diagrams,
into the propagator of the exchanged gluons, there is still a cancellation of
the Coulomb singularity, but the remainder grows like $n!$. These remainders
form an alternating series. which is Borel resummable. This series produces the
renormalon in the Borel transform at $u=- {1 \over 2}$. 

We proceed to  calculate  the Borel transform of the sum of diagrams in  
Fig.~4 in the full theory. The expression is identical to  
(\ref{Borel-G:2}), except that $\Gamma$ is
replaced by a momentum-dependent vertex $\Gamma(p-l, p+l)$:
\begin{eqnarray}
 && i {4 \pi } \; \left(\mu^2 \over e^C \right ) ^{u} \; 
 \int_l \; { 1 \over (-\l^2-i \epsilon)^{1 + u  } } \;
\nonumber \\ 
&& \hspace{0.7in} \times \,
 \bar{v}(p) \;\bigg[\;\gamma^\nu \; 
{ \not {\l} - \not \! { p} + m  \over {\l}^2 - 2\l \cdot p +i \epsilon } \;
T^a\, \Gamma (p-\l, p+\l) \,T^a\; 
{ \not {\l} + \not \! { p} + m  \over {\l}^2 + 2\l \cdot p +i \epsilon } \;
\gamma_\nu  \; \bigg] \;u(p)  \;.
\label{Borel-coulm:1} 
\end{eqnarray}
Since the Coulomb singularity comes from the region of
$l \ll m $, the $l$ terms in the numerator and in the vertex $\Gamma$ 
can be  neglected. Using $\not\! p u = m \, u$ and 
$\bar{v}\! \not\! p =-m\,\bar{v} $, 
 the above equation  can be reduced to  
\begin{eqnarray}
&&  -i \, {16 \pi  m^2 } \, \left(\mu^2 \over e^C \right ) ^{u}\; 
\,\bar{v}(p) \, T^a\,\Gamma (p,p) \,T^a \;u(p) \,
    \nonumber \\
&& \hspace{1in} \times \,
\int_l \, 
{ 1 \over (-\l^2 - i\epsilon)^{1 + u  } }  \;
{ 1\over  \l^2 -2m l_0  + i \epsilon } \,
{ 1\over  \l^2 +2m l_0  + i \epsilon } \;.
\label{Borel-coulm:2} 
\end{eqnarray}
If the annihilation vertex $\Gamma$ is a color-singlet, the factor 
$\bar{v} T^a \Gamma  T^a u$ reduces to $(4/3)  \bar{v}  \Gamma u$.
If $\Gamma$ is a color-octet vertex, the factor $4/3$ is replaced by $ - 1/6$. 

To evaluate this integral, 
we first integrate over $l_0$ by using the contour integral method. 
Closing the contour in the upper-half plane, 
the Coulomb singularity  
appears in the  contribution from 
the pole at 
$l^*_{0}= -m + \sqrt{m^2+{\bf l}^2 } +i \epsilon$. 
Keeping only this term,  (\ref{Borel-coulm:2}) reduces 
in the case of a color-singlet annihilation vertex to 
\begin{equation} 
 {16 \pi m^2 \over 3   } \; \left( {\mu^2 \over  e^C }\right ) ^{u} 
\; \bar{v}(p) \; \Gamma (p,p) \;u(p) \; 
\int  \; {d^3 {\rm\bf \l} \over (2 \pi)^3 }\; 
{ 1\over \left(  2m \sqrt{m^2+{\bf l}^2} -2 m^2 \right)^{2+u}    } \; 
{ 1\over \sqrt{m^2 + {\bf \l}^2 } } 
\;. 
\label{Borel-coulm:3} 
\end{equation} 
The first factor in the integrand reduces to $1/({\bf \l}^2)^{2+ u} $ 
for small ${\bf \l}$. It is easy to see that if (\ref{Borel-coulm:3}) 
is expanded as a power series in $ u $, all the coefficients have  a 
linear infrared divergence coming from the small $ {\rm\bf \l} $ region. 
This just corresponds to the Coulomb singularity. 

We next consider  the Borel transform of the sum of the diagrams
in Fig. 4 in NRQCD. Since there are no momentum scales in the diagrams and
since it must match (\ref{Borel-coulm:3}) at low momenta, the result must be 
\begin{equation}
%\widetilde{E}^{\Gamma} (u) \Gamma  \;=\; 
{16 \pi m \over 3 }  \; 
\left( {\mu^2 \over e^C} \right ) ^{u} \;   
\bar{v}(p) \; \Gamma(p,p) \;u(p) \;
\int  \; {d^3 {\rm\bf \l} \over (2 \pi)^3 }\;
{ 1\over  ({\bf l}^2)^{2+u}   } \;. 
\label{Borel-coulm:4} 
\end{equation}
The contribution of the Coulomb singularity to the 
short-distance coefficient $\widetilde{C}^{\Gamma} (t)$ is obtained by subtracting
(\ref{Borel-coulm:4}) from (\ref{Borel-coulm:3}) and removing the factor
$\bar{v} \Gamma u$.  The Coulomb singularity  cancels, so the power series  
 expansion of $\widetilde{C}^{\Gamma} (t)$  
has  finite coefficients.  However,  the integral in (\ref{Borel-coulm:4}) 
has a power ultraviolet divergence when $u > - {1 \over 2}$, and this produces
a pole in $\widetilde{C}^{\Gamma} (t)$ at $ u= -{1 \over 2} $. The pole can be
isolated by imposing a cutoff $|{\bf \l}| > \kappa $ on the integral: 
\begin{equation}
\widetilde{C}^{\Gamma} (t) \; \sim \; 
-{16 \pi m \over 3 }  \; \left( {\mu^2 \over e^C}\right ) ^{u} \;
\int _{|{\bf \l|} > \kappa} \; {d^3 {\bf \l} \over (2 \pi)^3 }\;
{ 1\over  ({\bf \l}^2)^{2+u}   } \;. 
\end{equation}
Evaluating the integral, we find that the residue of the pole is identical to
that in (\ref{C-u:-1/2}). 

It is easy to see that the renormalon at $ u= -{1 \over 2} $ and its residue
are gauge invariant. If we had used the gluon propagator  
(\ref{gauge}) to carry out the calculation in a general covariant gauge,
there would be additional terms in (\ref{Borel-coulm:1}) with $\gamma_\nu$
replaced by $\not \! l$. Making the decomposition  
$\not\! {\l}= (\not\!{p}-m ) - (\not\!{p} - \not\!{\l}-m ) $,
the first term vanishes due to the equation of motion for $u(p)$,  
while the second term cancels the quark propagator. 
This removes the pinch singularity, and therefore these terms have 
no pole at $ u= -{1 \over 2} $. 

The residue of the $u= -{1\over 2} $  
renormalon is universal for all color-singlet $S$-wave quarkonium annihilation
decays. The universality follows from the fact that in 
the low-momentum region of the
loop integral for the QCD amplitude (\ref{Borel-coulm:1}),  
the $c\bar{c}$ annihilation amplitude can be approximated by a
vertex $\Gamma(p,p)$ that is independent of the loop momentum and nonzero for
$S$-wave annihilation. 
The low-momentum regions of the QCD loop integral and the corresponding NRQCD
loop integral must match. But there are no momentum scales in the NRQCD loop
integral, and therefore its ultraviolet behavior is also fixed. The 
residue of the ultraviolet NRQCD renormalon at $u=-{1\over 2}$ 
depends only on whether the annihilation 
amplitude $\Gamma(p,p)$ is color-singlet or color-octet. 
Note that this renormalon is generated only 
by loop diagrams in which a gluon is exchanged between the $c$ and $\bar{c}$.  
Other loop diagrams, such as quark self-energy diagrams and the 
diagrams in which a gluon connects the $c$ or $\bar{c}$ line to the 
annihilation vertex, give no contributions  to the renormalon 
simply because there is no pinch singularity in the QCD diagrams.

%%%%%%%%%%%%%% 
\subsection { Cancellation of the $ u={ 1\over 2 } $ renormalon ambiguities } 
The Borel transforms $\widetilde{C}^{\Gamma}(t)$ in (\ref{C-V})-(\ref{C-P}) 
and (\ref{Borel-etacgg}) have an infrared renormalon at $u= +{1 \over 2}$. 
They can therefore be written in the form 
\begin{equation}
\widetilde{C}^{\Gamma}(t) \;=\; {a^\Gamma \over 1 - 2\beta_0 t } \,+\, 
\widetilde{A}^{\Gamma}(t) \;,
\label{C-u:1/2}
\end{equation}
where $\widetilde{A}^{\Gamma}(t)$ is analytic at $\beta_0t = {1\over 2}$  
and the residues of the poles are 
\begin{eqnarray}
a^V &\;=\;& -{11 \over 9 \pi }\,{\mu\over e^{C/2}\, m} \;, \label{a-V}\\
a^A &\;=\;& -{1 \over 3 \pi }\, {\mu\over e^{C/2}\, m} \;, \label{a-A}\\
a^T &\;=\;& -{5 \over 3 \pi }\, {\mu\over e^{C/2}\, m} \;, \label{a-T}\\
a^P &\;=\;&  {1 \over  \pi }\,  {\mu\over e^{C/2}\, m} \;, \label{a-P}\\
a^{\gamma\gamma} &\;=\;& {5 \over 9 \pi }\, {\mu \over e^{C/2}\,m}
\;.\label{a-gg}
\end{eqnarray}
Since the pole at $u={1\over 2}$ is located on the positive real axis, the inverse
Borel transform is not unique. Taking the principal value of the integral in
(\ref{borel-inverse}) is only one of the possible prescriptions.  
If ${A}^{\Gamma}(\alpha_s) $ is the inverse Borel transform of
$\widetilde{A}^{\Gamma}(t) $, the general expression for the 
inverse Borel transform of (\ref{C-u:1/2}) is 
\begin{equation}
C^\Gamma (\alpha_s) \;=\;  a^\Gamma\,  \alpha_s \;
{\rm\large P} \!\! \int^\infty_0 \, dx \, 
{ e^{- x} \over 1 - 2 \beta_0 \alpha_s \, x }
\,+\, {A}^\Gamma (\alpha_s ) \;
\,-\, K {\pi a^\Gamma \over \beta_0} \,e^{- 1/(2 \beta_0 \alpha_s)} \;,
\label{C-u:1/2:2}
\end{equation}
where  ${\rm\large P}\!\!\int$ denotes the principal value integral, 
$K$ is an arbitrary constant, and $\alpha_s= \alpha_s(\mu)$.  
The last term in (\ref{C-u:1/2:2})  represents the ambiguity in the inverse 
Borel transform, 
which is proportional to the residue of the pole in the integrand. 
Using the one-loop expression (\ref{alphas}) for the running coupling constant
$\alpha_s(\mu)$ and the fact that $a^\Gamma$ includes a factor of $\mu /m$, 
we see that the ambiguity in (\ref{C-u:1/2:2}) 
is proportional to $ \Lambda_{QCD} /m$. 
This ambiguity arises because summing  the series of loop diagrams with
bubble chains inserted into the gluon propagator  is equivalent to 
replacing  the strong coupling 
constant by a running coupling constant at the scale of the gluon momentum.
However, the perturbative running coupling constant has a Landau pole  
at the momentum  scale $\Lambda_{\rm\small QCD}$. 
This pole appears in the integration region and therefore gives rise to an
ambiguity in the loop integral. 
The renormalon ambiguity implies that
the short-distance coefficients defined by minimal subtraction necessarily have
some sensitivity to  nonperturbative effects 
involving the   energy scale $\Lambda_{\rm\small QCD}$. 

The infrared renormalons  give ambiguities in the short-distance 
coefficients in NRQCD factorization formulas. 
The decay rates for $J/\psi \to  e^+ e^-$ and $\eta_c \to \gamma\gamma$ and the
matrix elements in (\ref{Coe-A})-(\ref{Coe-P}) are physical quantities, 
so they cannot have  any ambiguities. 
The NRQCD factorization formulas simply represent the separation of scales. All
effects of momentum scales $m$ and larger are included in the short-distance coefficients, while
effects of lower scales, such as $mv$, $mv^2$, and $\Lambda_{\rm\small QCD}$,
are absorbed into the NRQCD matrix elements. Since 
infrared renormalons arise from small momentum scales,
 any ambiguities in the coefficients due to
infrared renormalons must be compensated by corresponding ambiguities in the
nonperturbative matrix elements. Thus it must be possible to consistently 
absorb the ambiguities
proportional to $K$ in (\ref{C-u:1/2:2}) for $\Gamma=V$, $A$, $T$, $P$ and
$\gamma\gamma$ into various NRQCD matrix elements. 

The ambiguities  in a coefficient from an infrared renormalon at $u=k/2$, where
$k$ is an integer, is proportional to $(\Lambda_{\rm\small QCD}/ m )^k$. Since
the NRQCD matrix elements involve several different scales, it is not
immediately obvious which matrix element an ambiguity proportional to
$\Lambda_{\rm\small QCD} ^k$ should be absorbed into. We will make a simple
hypothesis that is consistent with the velocity-scaling rules of NRQCD and
verify that the ambiguities due to the $u=+{1 \over 2}$ renormalon are
consistent with this hypothesis. Our hypothesis is that ambiguities proportional
to $\Lambda_{\rm\small QCD} ^k$ can be absorbed into 
NRQCD matrix elements that are suppressed by $v^{2k}$ or less. If our hypothesis
is correct, ambiguities due to the $u=+{1\over 2}$ renormalon should be absorbed
into the leading order matrix elements and those suppressed by $v^2$. In the
case of the $J/\psi$, these matrix elements are 
$    \langle 0 | \chi^\dagger \sigma^i \psi | J/\psi \rangle $ 
and 
$    \langle 0 | \chi^\dagger {\bf D}^2 \sigma^i \psi |J/\psi \rangle 
     \;.   $
In the case of  $\eta_c$, they are
$   \langle 0 | \chi^\dagger  \psi | \eta_c \rangle $ 
and 
$   \langle 0 | \chi^\dagger {\bf D}^2  \psi | \eta_c \rangle  $.
Because of the spin-symmetry relation $(\ref{spin-D2})$, there are only three
independent matrix elements that are suppressed by $v^2$ or less. It is
convenient to introduce the following shorthand notation for these three
independent matrix elements:  
\begin{eqnarray}
\langle  {\cal K}_{\psi} \rangle &\;\equiv \;&  
 \mbox{\boldmath $\epsilon$}\cdot 
 \langle 0 | \chi^\dagger \mbox{\boldmath $\sigma$}  \psi 
  |J/\psi(\mbox{\boldmath $\epsilon$})  \rangle,  \\
\langle  {\cal K}_{\eta} \rangle &\;\equiv \;&  
  \langle 0 | \chi^\dagger  \psi | \eta_c \rangle,  \\
 \langle {\cal K}_{D^2} \rangle &\;\equiv \;& {1\over m^2} \,  
   \langle 0 | \chi^\dagger {\bf D}^2  \psi | \eta_c \rangle 
    \;.
\end{eqnarray}

We first consider the cancellation of the renormalon ambiguities in 
the QCD matrix elements in
(\ref{Coe-V})-(\ref{Coe-P}). The coefficients of the matrix element
$ \langle {\cal K}_{D^2} \rangle $ in these factorization formulas are given in 
(\ref{A-expan:2}) and (\ref{Tensor-expan})-(\ref{Pseudo-expan}). The
coefficients $C^\Gamma$ are functions of $\beta_0 \alpha_s$ divided by 
$\beta_0$ and are therefore suppressed by $1/N_f$ in the large-$N_f$ limit. 
Since the ambiguities $\Delta C^\Gamma$ in the coefficients are suppressed 
by $1/N_f$, the ambiguities $\Delta \langle  {\cal K}_{\psi} \rangle $, 
$\Delta \langle  {\cal K}_{\eta} \rangle $, and 
$\Delta \langle {\cal K}_{D^2} \rangle $ in the matrix elements must also be
suppressed by $1/N_f$. The cancellation conditions for the renormalon  
ambiguities at leading order in $1/N_f$ are therefore linear relations:
\begin{eqnarray}
\Delta {C^V} \,\langle  {\cal K}_{\psi} \rangle 
&\;+\;& \Delta \langle {\cal K}_{\psi} \rangle
  \;+\;{1\over 6} \Delta \langle {\cal K}_{D^2} \rangle
  \;=\; 0  \;, 
\label{cancel-V} \\
\Delta {C^A} \,\langle  {\cal K}_{\eta} \rangle 
&\;+\;& \Delta \langle {\cal K}_{\eta} \rangle
  \;+\;{1\over 2} \Delta \langle {\cal K}_{D^2} \rangle
  \;=\; 0  \;,  
 \label{cancel-A}\\
\Delta {C^T} \,\langle  {\cal K}_{\psi} \rangle 
&\;+\;& \Delta \langle {\cal K}_{\psi} \rangle
  \;+\;{1\over 3} \Delta \langle {\cal K}_{D^2} \rangle
  \;=\; 0  \;, 
  \label{cancel-T}\\
\Delta {C^P} \,\langle  {\cal K}_{\eta} \rangle 
&\;+\;& \Delta \langle {\cal K}_{\eta} \rangle
  \;=\; 0  \;.
 \label{cancel-P}
\end{eqnarray}
	From (\ref{C-u:1/2:2}), the ambiguities in the coefficients are 
\begin{equation}
\Delta {C^\Gamma} \;=\; - K \,{\pi a^\Gamma \over \beta_0} \; 
{\Lambda_{\small\rm QCD} \over \mu}\;,
\label{D-C-G}
\end{equation}
with the coefficients $a^\Gamma$ given by (\ref{a-V})-(\ref{a-P}). The
cancellation conditions 
(\ref{cancel-V})-(\ref{cancel-P}) are four equations for the three unknown
ambiguities in the matrix elements, which provides a consistency check. The
solutions are  
\begin{eqnarray} 
\Delta \langle{\cal K}_\psi\rangle &\;=\;& 
    - K \, {7 \over 9 \,\beta_0} \, 
     {\Lambda_{\rm\small QCD}  \over e^{C/2} \,  m  } \, 
     \langle{\cal K}_\psi\rangle\; ,
\label{Delta:K-psi}\\ 
\Delta \langle{\cal K}_\eta \rangle &\;=\;& 
     K \, {1 \over \beta_0}  \, 
     {\Lambda_{\rm\small QCD}  \over e^{C/2} \, m  } \, 
     \langle{\cal K}_\eta \rangle\; ,
\label{Delta:K-eta}\\ 
\Delta \langle{\cal K}_{D^2} \rangle &\;=\;& 
    - K \, {8 \over 3 \,\beta_0}  \, 
     {\Lambda_{\rm\small QCD}  \over e^{C/2} \, m  } \, 
     \langle{\cal K}_\eta \rangle\; .
\label{Delta:K-D2}
\end{eqnarray}
These results can be verified by perturbative calculations of the large-$N_f$
limits of the Borel transforms of the NRQCD matrix elements. 

Now we examine the renormalon cancellations for the process $\eta_c \to
\gamma\gamma$. From (\ref{expan-A-etac:4}) and (\ref{C^gg:0}), the factorization
formula for the $T$-matrix element can be written in the form
\begin{eqnarray} 
 {\cal T}_{\eta_c \to \gamma\gamma} &\;=\;& - i  \;
 {2 Q^2  e^2   \over  M_{\eta_c}   } \; 
 \mbox{\boldmath $\epsilon$}_1 \times \mbox{\boldmath $\epsilon$}_2 
 \cdot \hat {\bf r}\;
%  \nonumber \\ && \;   \times 
    \Bigg[ \; \left(\,1+ C^{\gamma\gamma}\,\right)\, 
   \langle {\cal K}_\eta \rangle 
     \; + \; {1 \over 6 } \,  
   \langle {\cal K}_{D^2}\rangle  
 \; \Bigg] \;.
\label{etac:factor}
\end{eqnarray}
We have used the identity (\ref{Metac-m}) to eliminate the pole mass 
$m_{\rm pole}$ in  the prefactor in favor  of the meson mass $M_{\eta_c}$. 
The cancellation condition for the ambiguities from the renormalon at
$u={1\over 2}$ is
\begin{eqnarray} 
   \Delta C^{\gamma\gamma} \,  
 \langle   {\cal K}_\eta \rangle \;
 +\;\Delta \langle {\cal K}_\eta \rangle\;
 +\;  {1 \over 6}\, 
   \Delta \langle {\cal K}_{D^2} \rangle  &\;=\;& 0\;.
\label{cancel:gg}
\end{eqnarray}
Using the value of $a^{\gamma\gamma}$ in (\ref{a-gg}) and the values of   
$\Delta \langle {\cal K}_\eta \rangle$ and 
$\Delta \langle {\cal K}_{D^2} \rangle$ 
given in  (\ref{Delta:K-eta}) and (\ref{Delta:K-D2}),  
we see that (\ref{cancel:gg}) is satisfied.

We can use the results given above to verify the cancellation of renormalon
ambiguities in the Gremm-Kapustin identity, which is given at tree level in 
(\ref{Gremm-Kapustin}). The correct generalization to higher order in
$\alpha_s$  is to replace $m$ by $m_{\rm pole}$. The analogous identity for
the matrix elements of $\eta_c$ is 
\begin{equation}
   \langle 0 | \chi^\dagger {\bf D}^2
   \psi |\eta_c \rangle  
  \;=\; -m_{\rm pole} (M_{\eta_c} - 2 m_{\rm pole})\,  
   \langle 0 | \chi^\dagger  \psi | \eta_c \rangle \,
   \left(\, 1 + O(v^2) \, \right) \;.
\label{Gremm-Kapustin:etac}
\end{equation}
	From the expression (\ref{ren-M}) for the Borel transform of 
$m_{\rm\small pole}$, we find that the ambiguity due to the renormalon 
at $u=+{1\over 2}$ is 
\begin{equation}
\Delta m_{\rm\small pole} \;=\; - K \, {4 \over 3 \,\beta_0} \, e^{-C/2} \, 
     {\Lambda_{\rm\small QCD}  } \; .
\label{Delta:m_pole}
\end{equation}
The leading contribution to the ambiguity on the right side of
(\ref{Gremm-Kapustin:etac}) comes from the factor 
$M_{\eta_c} - 2 m_{\rm pole}$. The ambiguities from the  factors on the right
side of  (\ref{Gremm-Kapustin:etac}) are suppressed by $v^2$ from the factor 
$M_{\eta_c} - 2 m_{\rm pole}$. Thus the renormalon cancellation condition is
\begin{equation}
\Delta \langle {\cal K}_{D^2} \rangle \;=\; 2 \, 
{\Delta m_{\rm pole} \over m }\, 
 \langle {\cal K}_{\eta} \rangle \;. 
\end{equation}
Using (\ref{Delta:K-D2}) and (\ref{Delta:m_pole}), we see that this is
satisfied. 

%%%%%%%%%%%%%%%
\subsection{Leading Renormalons for $J/\psi \to \gamma\gamma\gamma$ } 
The one-loop QCD correction to the short-distance coefficient for 
the process $J/\psi \to \gamma\gamma\gamma $ has been  calculated
by Mackenzie and Lepage\cite{ml}. The correction to the decay rate is a
multiplicative factor $1-12.6 \alpha_s/\pi$, which gives a 
correction of about $-140\%$  if
we take the coupling constant to be $\alpha_s(m) \approx 0.35$. 
Since it is  the leading order correction, the coefficient of $\alpha_s$
is independent of the renormalization scale and the renormalization 
scheme. The enormous radiative correction raises  questions  
about  the origin of the large coefficient of $\alpha_s$  
and about the  convergence of the perturbation series. 
Some insights can be gained  by studying 
the renormalon singularities for this process. As shown above, 
the large order behavior of the short-distance coefficient 
is governed by the $u=\pm {1\over 2}$ renormalons. 
We know the residue of the  $u=-{1\over 2}$ renormalon in the large 
$N_f$ limit, since 
it is universal for all $S$-wave annihilation decay processes. 
The residue of the $u= {1\over 2}$ renormalon  
can be determined by requiring the cancellation of the renormalon 
ambiguities in the decay width.  
Knowing the residues of the leading renormalons, 
we can estimate the  asymptotic behavior 
of the perturbation series.  

The decay width for $J/\psi \to \gamma\gamma\gamma$, with the coefficient of
the first relativistic correction calculated to leading order, is
\begin{eqnarray} 
{\Gamma}(J/\psi \to \gamma\gamma\gamma) &\;=\;& 
 {16(\pi^2-9)Q^6\alpha^3 \over 9 M_{J/\psi}^3 } \; 
 \bigg| \; \left(\,1+C^{3\gamma}(\alpha_s)\,\right)\, 
   \langle 0 | \chi^\dagger \mbox{\boldmath $\sigma$} \psi | J/\psi 
   \rangle  
   \nonumber \\
   && \hspace{1in} 
 \,+\, { 7\pi^2 -24 \over 24 (\pi^2-9) } \,{1\over m^2} \, 
   \langle 0 | \chi^\dagger {\bf D}^2 \mbox{\boldmath $\sigma$} 
   \psi |J/\psi \rangle  \, 
  \bigg|^2  \;.
\label{Wid-psiggg}
\end{eqnarray} 
The coefficient of the matrix element 
$\langle 0 | \chi^\dagger {\bf D}^2 \mbox{\boldmath 
$\sigma$} \psi |J/\psi \rangle $ was first calculated by 
Keung and Muzinich in numerical form\cite{keung} and by Schuler in 
analytic form\cite{schuler}. 
The condition for the cancellation of the $u={1 \over 2}$ 
renormalon ambiguities in  
${\Gamma}(J/\psi \to \gamma \gamma\gamma) $ reads: 
\begin{eqnarray}
\Delta C^{3\gamma}  \,\langle  {\cal K}_{\psi} \rangle 
&\;+\;& \Delta \langle {\cal K}_{\psi} \rangle
  \;+\; { 7\pi^2 -24 \over 24 (\pi^2-9) } 
  \Delta \langle {\cal K}_{D^2} \rangle 
  \;=\; 0  \;.
\label{cancel-ggg} 
\end{eqnarray}
Using the values of $\Delta \langle {\cal K}_{\psi} \rangle$ 
and $\Delta \langle {\cal K}_{D^2} \rangle $ determined in
(\ref{Delta:K-psi}) and (\ref{Delta:K-D2}), we obtain  
\begin{eqnarray} 
\Delta C^{3\gamma} &\; = \;&
 K \, {14 \pi^2 -87 \over 9(\pi^2-9)  \,\beta_0}  \, 
     {\Lambda_{\rm\small QCD}  \over e^{C/2} m }
 \;.
\label{Delta-C-ggg} 
\end{eqnarray}
Comparing with (\ref{D-C-G}), we can read off the residue $a^{3 \gamma}$ of the
$u={1\over 2}$ renormalon, which is defined by (\ref{C-u:1/2}): 
\begin{eqnarray}
a^{3\gamma} &\;=\;&
 -{14 \pi^2 -87 \over 9(\pi^2-9)  \pi} \;  {\mu \over e^{C/2} m} 
 \;.
\label{a-ggg} 
\end{eqnarray}
This residue is rather large. It differs from the residues $a^V$ and
$a^{\gamma\gamma}$ for the decays $J/\psi \to e^+e^-$
and $\eta_c \to \gamma \gamma$ by factors of $5.4$ and $-11.8$, respectively.
The large one-loop radiative correction may be related to the large residue of
the $u= {1\over 2}$ renormalon.

%%%%%%%%%%%%%%%%%%%%%%%%%%%%%%%%%%%%%%%%%%%%%%%%%%%%
\section{ Estimates of Higher Order Radiative Corrections}
One of the problems encountered in phenomenological applications 
of the NRQCD factorization formalism is that the
one-loop corrections to the short-distance coefficients for several decay rates
are uncomfortably large. This raises questions about the convergence of the QCD
perturbation series.  The calculations of 
the short-distance coefficients in the large-$N_f$ limit can 
shed some light on the origin of the large radiative corrections.  
They  can also be used to provide
simple estimates of higher order radiative corrections. It has been found 
empirically that if we take the coefficients of the perturbation series in the
large-$N_f$ limit and make the shift $N_f \to N_f - 33/2$, we obtain reasonable
approximations to the higher order coefficients in cases where they are known
exactly\cite{BBB}. This prescription, which has been called ``naive nonabelianization'',
puts the Borel singularities at the correct points on the $t$-axis.
We will use this prescription to give simple estimates of higher order corrections
in the perturbation series for $C^V(\alpha_s)$,  
$C^{\gamma\gamma}(\alpha_s)$, and $C^{3\gamma}(\alpha_s)$. 

The Borel transform $\widetilde{C}^\Gamma(t)$ for the short-distance
coefficients in the decay rate for $J/\psi \to e^+e^-$ and $\eta_c \to \gamma
\gamma$ are given in (\ref{C-V}) and (\ref{Borel-etacgg}). 
To obtain the perturbation expansion for $C^\Gamma(\alpha_s)$ in the 
large-$N_f$ limit, we expand $\widetilde{C}^\Gamma (t)$ as a power series in
$u$, and then invert the Borel transform by substituting 
$t^{n-1} \to (n-1)! \, \alpha_s^n$.  
We set $\beta_0 = (33-2N_f)/ 12 \pi$ and take $N_f=3$ for charmonium decays. 
We choose the $\overline{\rm MS}$ scheme by setting $C=-5/3$ and
we take the scale of the running coupling constant to be $\mu = m $. The
resulting perturbation series are 
\begin{eqnarray}
C^V(\alpha_s) &\;=\;&  -{ 2.7} \,{a } \,  \,-\, 5.5   \,a^2 \,
 -\,150\, a^3 \, -\,870\, a^4 \,
% \nonumber \\
% && \hspace{1in}
  -\,38000\, a^5 \, -\,360000\, a^6 \, + \, \cdots \;,
 \label{C-V-expan}\\ 
C^{\gamma\gamma} (\alpha_s) &\;=\;& 
-{ 1.7} \,a \,  \,+\, 5.1   \,a^2 \, -\,25\, a^3 \, +\,1100\, a^4 \,
% \nonumber \\
% && \hspace{1in}
 -\,900\, a^5 \, +\,520000\, a^6 \, + \, \cdots \;.
 \label{C-gg-expan}
\end{eqnarray}
where $a=\alpha_s(m) /\pi$.
We have kept only two significant figures in each of the coefficients.

The behavior of the perturbation series (\ref{C-V-expan}) and 
(\ref{C-gg-expan}) can be understood by examining the contributions from 
 the renormalon 
singularities at $u={1\over 2}$ and $u=-{1\over 2}$. Asymptotically, these
renormalons dominate the coefficients in the perturbation series, 
because the contribution to the coefficient of $\alpha_s^n$ from 
a higher renormalon at $u=k/2$ is 
suppressed  by $k^{-n}$. The contributions to the perturbation series
(\ref{C-V-expan}) and (\ref{C-gg-expan}) from the renormalons 
at $u=\pm {1 \over 2}$   are  
\begin{eqnarray}
C^V (\alpha_s) &\;\sim\;& 
{1 \over 2 \pi \beta_0} \; \sum_{n} \;
\left[ \;-{11\over 9 } \,{\mu \over e^{C/2}\, m} 
\, +\,  (-1)^n \, {8\over 3} \,{ e^{C/2}\, m\over\mu }  \; \right] 
 \; (n-1) !\, [2 \beta_0\alpha_s(\mu)]^{n}  \;, 
\label{C-V-asymp} \\ 
C^{\gamma\gamma} (\alpha_s) &\;\sim\;& 
{1 \over 2 \pi \beta_0} \; \sum_{n} \;
\left[ \;{5\over 9 } \,{\mu \over e^{C/2}\, m} 
\, +\,  (-1)^n \, {8\over 3 } \,{ e^{C/2}\, m\over\mu }  \; \right] 
 \; (n-1) !\, [2 \beta_0\alpha_s(\mu)]^{n}  \;. 
\label{C-gg-asymp} 
\end{eqnarray}
The alternating  series in (\ref{C-V-asymp}) and  (\ref{C-gg-asymp})
are identical, since the $u=-{1\over 2}$ renormalon is  universal 
for all $S$-wave annihilation decay processes. 
The same-sign series in (\ref{C-V-asymp}) and  (\ref{C-gg-asymp})
are proportional to the residues of the $u=+{1\over 2}$ renormalon. 
The relative importance  of the two renormalons 
depends on the renormalization  scale $\mu$. The  $u=-{1\over 2}$ 
renormalon  dominates if $\mu \ll m$ and the $u=+{1\over 2}$ 
renormalon dominates if  
$\mu\gg m$. The choice of $\mu$ that gives equal contributions from the two
renormalons is $\mu=1.48 e^{C/2} m$ for  (\ref{C-V-asymp}) and 
$\mu=2.19 e^{C/2} m$ for  (\ref{C-gg-asymp}). In the $\overline{\rm MS}$
scheme, these scale are $0.64 m$ and $0.95 m$. These values suggest that
$\mu=m$ is a reasonable value for the renormalization scale  in the 
$\overline{\rm MS}$ scheme. For comparison, the corresponding BLM scales are
$\mu=0.63m$ and $\mu=1.95m$, respectively. 
 
   In the $\overline{\rm MS}$ scheme ($C= - {5 \over 3 }$) with the scale 
$\mu=m$, the  first few terms in the series (\ref{C-V-asymp}) and 
(\ref{C-gg-asymp}) are
\begin{eqnarray}
C^V(\alpha_s) &\; \sim \;&  -{ 4.0} \,{a } \,  \,-\, 7.4   \,a^2 \,
 -\,160\, a^3 \, -\,900\, a^4 \,
% \nonumber \\
% && \hspace{1in}
  -\,39000\, a^5 \, -\,370000\, a^6 \, + \, \cdots \;,
 \label{C-V-expan:1/2}\\ 
C^{\gamma\gamma} (\alpha_s) &\;\sim \;& 
{ 0.12} \,a \,  \,+\, 11  \,a^2 \, +\,4.8\, a^3 \, +\,1300\, a^4 \,
% \nonumber \\
% && \hspace{1in}
 +\,1200\, a^5 \, +\,540000\, a^6 \, + \, \cdots \;.
 \label{C-gg-expan:1/2}
\end{eqnarray}
Comparing (\ref{C-V-expan}) and (\ref{C-V-expan:1/2}), we see that the leading
renormalons begin to dominate the large-$N_f$ coefficients of $C^V$ 
at the $\alpha_s^3$
term. Comparing (\ref{C-gg-expan}) and (\ref{C-gg-expan:1/2}), we see that the
leading renormalons dominate the even coefficients of $C^{\gamma\gamma}$ 
beginning at 
the $\alpha_s^4$ term, but they do not dominate  the odd coefficients 
until much higher order in $\alpha_s$. The reason for this is that  
the coefficient in square brackets 
in (\ref{C-gg-asymp}) is $1.28+(-1)^n 1.16$. 
The asymptotic coefficients for 
$C^{\gamma\gamma}$ are therefore positive and grow  like $(n-1)! \beta_0 ^{n-1}$. 
However,  for odd $n$, there is a near 
cancellation between the contributions of the  renormalons at 
$u=\pm {1\over 2}$ that delays 
the onset of the asymptotic behavior. At preasymptotic  values of $n$, the
odd coefficients may be dominated by the next higher renormalons at $u=\pm1$. 

We now consider the higher order radiative corrections to 
the decay rate for 
$J/\psi \to \gamma\gamma\gamma$. The large-$N_f$ limit  
of the short-distance coefficient 
$C^{3 \gamma}(\alpha_s)$  has not yet been calculated. Only the first
term in the perturbation series is known\cite{ml}:
\begin{equation}
C^{3 \gamma}(\alpha_s) \;=\; -6.3 a \,+\, \cdots \;.
\label{C-ggg-leading}
\end{equation}
However we have determined the residues of the $u=\pm { 1\over 2}$ renormalons,
and we can use that information to deduce the asymptotic behavior of the
coefficients in the large-$N_f$ limit.  
 The contributions from the   $u=-{1 \over 2}$ and  
$u={1 \over 2}$ renormalons are
\begin{eqnarray}
C^ {3 \gamma}(\alpha_s) &\;\sim\;& 
{1 \over 2 \pi \beta_0} \; \sum_{n} \;
\left[ \;-{ 14 \pi^2 -87 \over 9(\pi^2-9) } \,{\mu \over e^{C/2}\, m} 
\, +\,  (-1)^n \, {8\over 3} \,{ e^{C/2}\, m\over\mu }  \; \right] 
 \; (n-1) !\, [2 \beta_0\alpha_s(\mu)]^{n}  \;. 
\label{C-ggg-asymp}
\end{eqnarray}
In the $\overline{\rm MS}$ scheme ($C=-5/3$) with $\mu=m$,  
 the first few terms in the series are
\begin{eqnarray}
C^{3\gamma} (\alpha_s) &\; \sim\;&  
-{ 16 } \,a\, 
 \,-\, 62   \,a^2 \,
 -\,660\, a^3 \,
 -\,7600\, a^4 \,
% \nonumber \\
% && \hspace{1in}
  -\,160000\, a^5     -\,3100000\, a^6 \,
 + \, \cdots \;,
 \label{C-ggg-expan} 
\end{eqnarray}
where $a=\alpha_s(m) /\pi$.  
The asymptotic coefficients are almost an order of magnitude large than those in
(\ref{C-V-expan:1/2}) and (\ref{C-gg-expan:1/2}). The large coefficients are caused
by the large residue of the $u=+ {1\over 2}$ renormalon. 

One can attribute the large residue of the $u=+ {1\over 2}$ renormalon in
$C^{3\gamma}(\alpha_s)$ to a poor choice of the renormalization scale 
$\mu$. The
contribution from the $u=+{1\over 2}$ renormalon can be decreased at the expense
of increasing the contribution from the $u=- {1\over 2}$ renormalon. The best
balance between the two sources of $n!$ growth is to choose $\mu$ so that the two
renormalons contribute equally to the asymptotic  coefficients.
This prescription gives the rather small value $\mu= 0.28 m$ for the scale.  
With this choice of scale, the even coefficients in (\ref{C-ggg-expan}) vanish
and the odd coefficients are decreased in magnitude by about a factor of $2$.
This scale is rather close to the BLM scale for the closely related process
$J/\psi \to ggg$, which is $\mu=0.31m$. This suggests that this may indeed be
an  appropriate choice for $\mu$ in the $\overline{\rm MS}$ scheme.

We now consider briefly the phenomenological implications of our calculations
for charmonium decays, where the appropriate coupling constant is $\alpha_s(m)
\approx 0.35$. The first terms in (\ref{C-V-expan}), (\ref{C-gg-expan}), 
and  (\ref{C-ggg-leading})  give the complete 
order-$\alpha_s$ corrections to the decay amplitudes. They
represent radiative corrections of $-30\%$ for $J/\psi \to e^+ e^-$, 
$-19\%$ for $\eta_c \to \gamma\gamma$, and $-70 \%$ for $J/\psi \to
\gamma\gamma\gamma$. The order-$\alpha_s^2$ correction is known only in the
large-$N_f$ limit for $J/\psi \to e^+e^-$ and $\eta_c \to \gamma\gamma$, and
it gives corrections of $-7\%$ and $+6\%$. For $J/\psi \to
\gamma\gamma\gamma$, we only know the contribution  to the 
order-$\alpha_s^2$ correction 
from the renormalons at $u=\pm {1\over 2}$, which give a
correction of $-70\%$. We conclude that the perturbation series for the
amplitudes for the decays  $J/\psi
\to e^+ e^-$ and $\eta_c \to \gamma\gamma$ appear to be well-behaved through
next-to-next-to-leading order in $\alpha_s$. However, the perturbation series
for $J/\psi \to \gamma\gamma\gamma$ appears to be hopelessly divergent. One
might be able to decrease the order-$\alpha_s^2$ correction by changing the
renormalization scale to $\mu \approx m/3$, but only at the cost of increasing
the order-$\alpha_s$ correction to well over $100\%$. 

%%%%%%%%%%%%%%%%%%%%%%%%%%%%%%%%%%%%%%%%%%%%%%%%% 
\section{conclusion} 
In this paper, we studied the large-order asymptotic behavior of the
perturbation series 
for short-distance coefficients in the NRQCD factorization 
formalism. We calculated the Borel transforms of the short-distance 
coefficients for the electromagnetic decays $J/\psi \to e^+ e^-$ and 
$\eta_c \to \gamma\gamma$ in the large-$N_f$ limit. 
We found that there exists a universal resummable renormalon 
associated with the cancellation of the Coulomb singularity  in the 
short-distance coefficients. We verified   that the  ambiguities 
in the short-distance coefficients due to the first  infrared
renormalon  are canceled by ambiguities in the long-distance matrix elements
that contribute through relative order $v^2$. 
We used our calculation to make estimates of the higher order radiative
corrections to the decay rates for $J/\psi \to e^+ e^-$, 
$\eta_c \to \gamma\gamma$, and  $J/\psi \to \gamma \gamma\gamma$. 
This work represents a first step toward studying the effects 
of renormalons in the NRQCD factorization formalism.
We hope that further work in this direction can lead to an understanding of
how to deal with the large radiative corrections encountered in some decay
rates, such as $J/\psi \to \gamma\gamma\gamma$ and $J/\psi \to g g g $.

\acknowledgements

This work was supported in part by the U.S.
Department of Energy, Division of High Energy Physics, under 
Grant DE-FG02-91-ER40684. We acknowledge valuable discussions with G.T.
Bodwin.

\vfill \eject
%%%%%%%%%%%%%%%%%%% FIGURE CAPTIONS %%%%%%%%%%%%%%%%%%%%%%%%%

\begin{figure}
{ Fig.~1. 
 Tree-level diagram for $c \bar c \to e^+ e^-$.}
\end{figure}

\begin{figure}
{Fig.~2. Tree-level diagrams for $c \bar c \to \gamma \gamma$.}
\end{figure}

\begin{figure}
{Fig.~3.  One-loop diagram for the heavy-quark self-energy,
with a chain of light quark bubbles inserted into the gluon propagator.}
\end{figure}

\begin{figure}
{Fig.~4.  One-loop vertex correction for $c\bar{c} \to e^+ e^-$,
with a chain of light quark bubbles inserted into the gluon propagator.}
\end{figure}

\begin{figure}
{Fig.~5. One-loop diagrams for $c \bar c \to \gamma \gamma$,
with a chain of light quark bubbles inserted into the gluon propagator.
For each diagram, there is a corresponding diagram with the photon lines 
interchanged.  The blob in Figure 5e represents the mass counterterm.}
\end{figure}

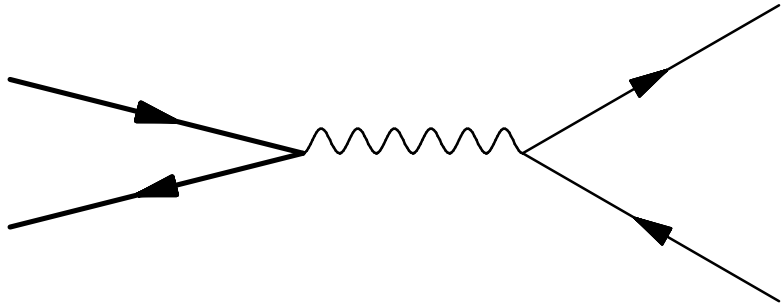
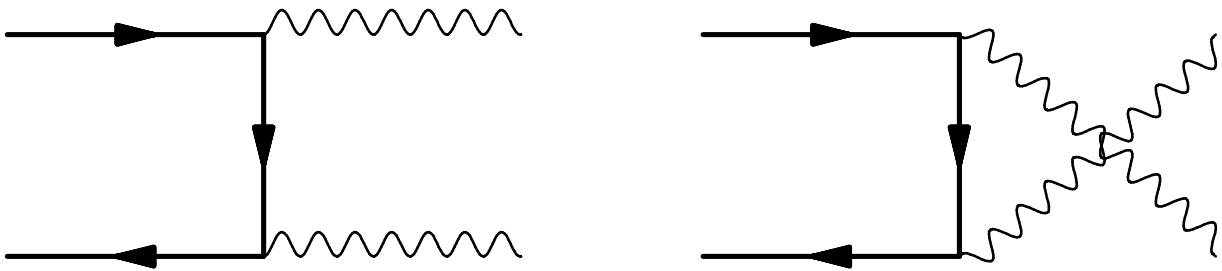
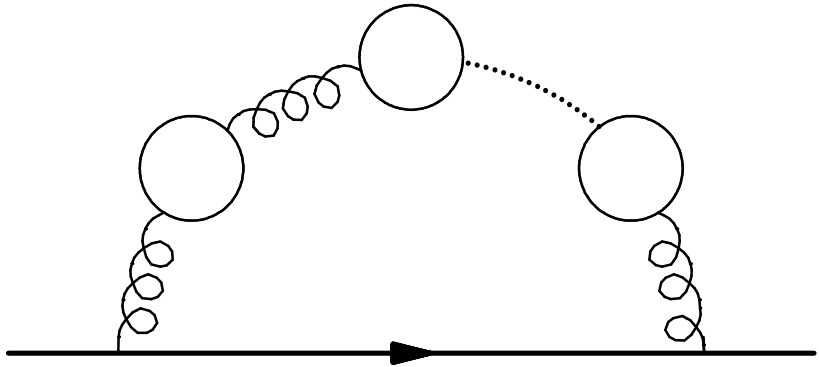
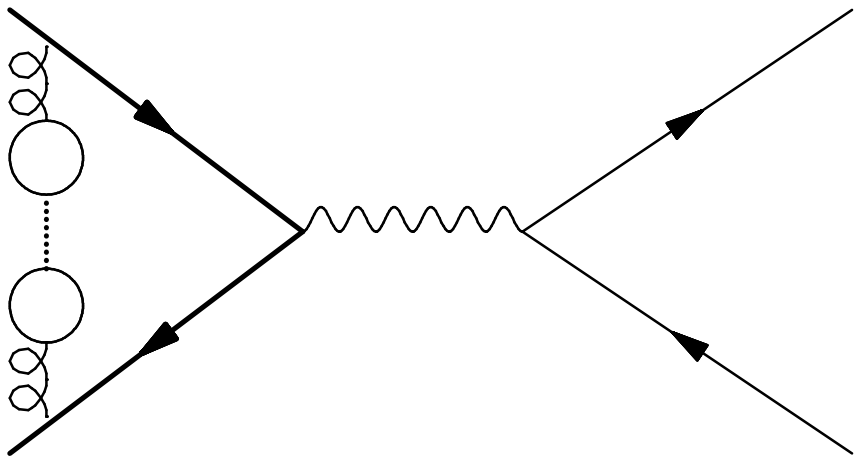
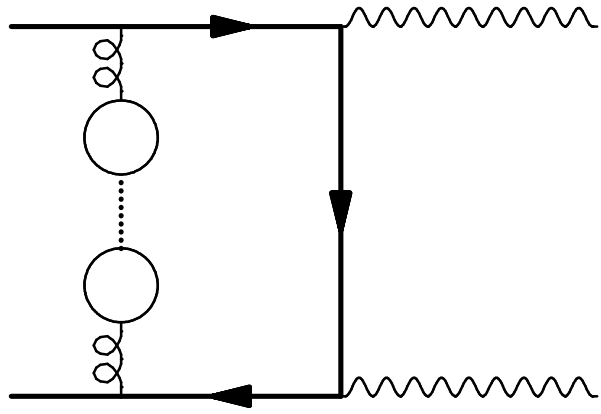
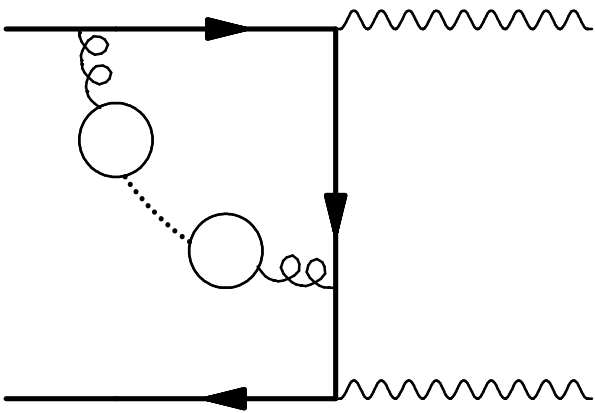
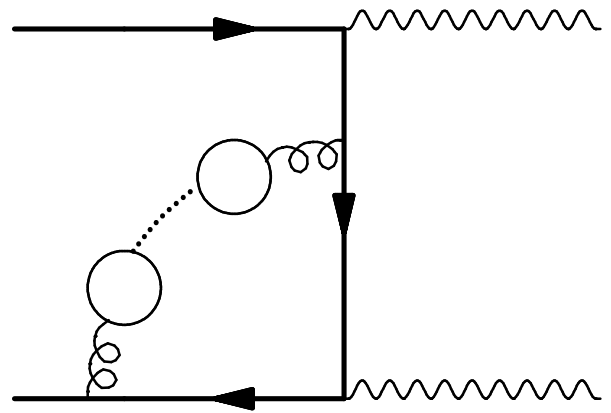
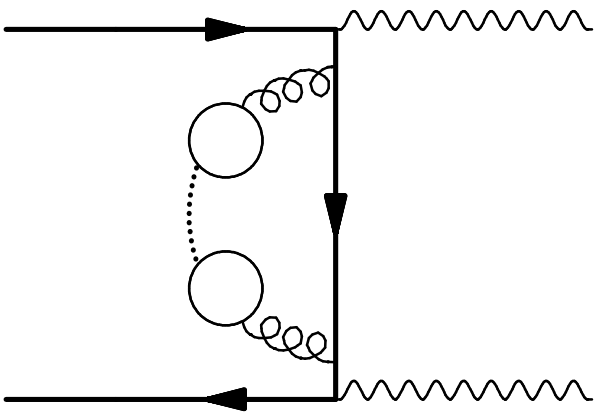
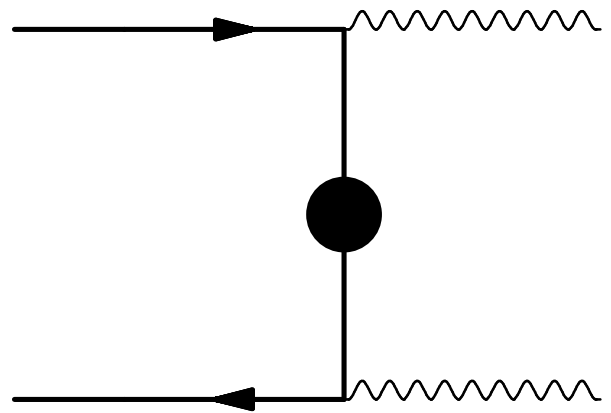
\end{document}